\documentclass[11pt]{article}%
\usepackage{amsmath}
\usepackage{graphicx}%
\usepackage{amsfonts}%
\usepackage{amssymb}
\newcommand{\eqnb}{\begin{equation}}
\newcommand{\eqne}{\end{equation}}

\begin{document}

\title{Reward Processes and Performance Simulation in Supermarket Models with
Different Servers}
\author{Quan-Lin Li and Feifei Yang\\
School of Economics and Management Sciences \\
Yanshan University, Qinhuangdao 066004, P.R. China\\
Na Li\\Department of Industrial Engineering and Management\\
Shanghai Jiaotong University, China}
\date{(Published in {\bf International Journal of Simulation and Process Modelling}, 2016, Special Issue: Simulation Modelling and Optimisation of Large-scale Systems, Pages 1--15)}
\maketitle

\begin{abstract}
Supermarket models with different servers become a key in modeling resource
management of stochastic networks, such as, computer networks, manufacturing
systems, transportation networks and healthcare systems. While the different
servers always make analysis of such a supermarket model more interesting,
difficult and challenging. This paper provides a novel method for analyzing
the supermarket models with different servers through a multi-dimensional
continuous-time Markov reward process. Firstly, some utility functions are
constructed for the routine selection mechanism according to the queue
lengths, the service rates, and the probability of individual preference.
Secondly, applying the state jump points of the continuous-time Markov reward
process, some segmented stochastic integrals of the random reward function are
established by means of an event-driven technique. Based on this, the mean of
the random reward function in a finite time interval is computed, and the mean
of the discounted random reward function in an infinite time interval can also
be calculated. Finally, some simulation experiments are given to indicate how
the expected queue length of each server depends on some key parameters of
this supermarket model.


\end{abstract}

%\author{Quan-Lin Li and Feifei Yang\\School of Economics and Management Sciences \\Yanshan University, Qinhuangdao 066004, P.R. China\\Na Li\\Department of Industrial Engineering and Management\\Shanghai Jiaotong University, China}

\section{Introduction}

Randomized load balancing, where a job is assigned to a server from a small
subset of randomly chosen servers, is very simple to implement, and can
surprisingly deliver better performance (for example, reducing collisions,
waiting times, and backlogs) in a number of applications, such as, data
centers, capacity allocation, hash tables, distributed memory machines, path
selection, and task scheduling. The supermarket model is a dynamic randomized
load balancing method, and its original idea may be inspired by operation
mechanism of supermarket checkout in a large supermarket. Because the
supermarket model has simple operations, quick response, dynamical real-time
management, and many other advantages, it has been regarded as one of the most
effective technologies in the study of large-scale stochastic networks with
resource management and task scheduling.

During the last two decades considerable attention has been paid to studying
the supermarket models through queueing theory as well as Markov processes.
Since a simple supermarket model was discussed by Mitzenmacher \cite{Mit:1996}%
, Vvedenskaya et al. \cite{Vve:1996} and Turner \cite{Tur:1996, Tur:1998},
subsequent papers have been published on this theme, among which, see,
Vvedenskaya and Suhov \cite{Vve:1997}, Mitzenmacher et al. \cite{Mit:2001},
Graham \cite{Gra:2000b, Gra:2004}, Luczak and Norris \cite{LucN:2005}, Luczak
and McDiarmid \cite{Luc:2006, Luc:2007}, Brightwell and Luczak \cite{Bri:2012}%
, Bramson et al. \cite{Bra:2010, Bra:2012, Bra:2013}, Li and Lui
\cite{LiL:2010, LiL:2014}, Li et al. \cite{Li:2011, Li:2013, Li:2015a,
Li:2015b} and Li \cite{Li:2011a, Li:2014}. For the fast Jackson networks (or
the supermarket networks), readers may refer to Martin and Suhov
\cite{Mar:1999}, Martin \cite{Mar:2001} and Suhov and Vvedenskaya
\cite{Suh:2002}. On the other hand, Janssen \cite{Jan:2011} applied the
discrete-time Markov reward processes as well as the discrete-time Markov
decision processes to the study of supermarket models with $N$ identical servers.
The stability of more general supermarket models was discussed by Foss and
Chernova \cite{Fos:1998}, Bramson \cite{Bra:2011} and MacPhee et al.
\cite{Mac:2012}.

There are some successful research on various Markov reward processes,
important examples include Reibman et al. \cite{Rei:1989}, Ciardo et al.
\cite{Cia:1990}, Qureshi and Sanders \cite{Qur:1994}, Telek et al.
\cite{Tel:1998}, de Souza e Silva and Gail \cite{de:1998}, Telek and R\'{a}cz
\cite{Tel:1999}, Telek et al. \cite{Tel:2004}, Li and Cao \cite{Li:2004},
Stefanov \cite{Ste:2006}, Stenberg et al. \cite{Ste:2007}, and two books by
Cao \cite{Cao:2007} and Li \cite{Li:2010}.

Little work has been done on analysis of the supermarket models with different
servers, which is more difficult and challenging due to high complexity and
percipient subjectivity of designing a fair routine selection mechanism with
respect to the different servers. Specifically, a practical understanding can
indicate that such a routine selection mechanism may depend on the queue
lengths, on the service rates, on the probability of individual preference and
so forth. Janssen \cite{Jan:2011} described a simple intuitive outline of
discussing the supermarket model with different servers, and demonstrated that
analysis of the supermarket model with different servers will be an
interesting and difficult topic in the future research. Based on this, Li et
al. \cite{Li:2015b} provided a birth-death reward process for the supermarket
model with different servers, and established a system of functional reward
equations which can be solved by a value iterative algorithm. It is worth
noting that this paper uses a more general Markov reward process to set up the
segmented stochastic integrals of the random reward function in the
supermarket model with different servers\ by means of an event-driven
technique, which is shown to be useful for performance simulation of a more
general large-scale stochastic system. In addition to this, we would like to
remark two key points: (1) Although the mean-field theory is an effective
method in the study of supermarket models with the same servers (e.g., see
Vvedenskaya et al. \cite{Vve:1996}, Li et al. \cite{Li:2013} and Li and Lui
\cite{LiL:2014}), the complicated routine selection mechanism with respect to
the different servers makes setting up the systems of mean-field equations
more difficult. To our best knowledge, up to now no paper has applied the
mean-field theory to the study of supermarket models with different servers.
(2) The generating functions are always classical and effective for
performance evaluation of many practical stochastic systems, but they are not
convenient to deal with a multi-dimensional problem, and are also very
difficult to analyze a system of nonlinear equations.

The main contributions of this paper are twofold. The first one is to describe
a supermarket model with different servers, in which the arrival and service
processes are given in a detailed discussion, and the reward value at each
state is chosen from some practical points of view. We show that the arrival
process of this supermarket model is very complicated due to a routine
selection mechanism that depends on the queue lengths, on the service rates,
on the probability of individual preference and so forth. Also, it is seen
that the routine selection mechanism is very different from that in the
supermarket model with same servers, where our construction of this routine
selection mechanism is based on the utility functions so that the subjective
behavior of customers is also covered in the routine selection mechanism. The
second one is to set up a multi-dimensional continuous-time Markov reward
process, and provide a segmented stochastic integral for expressing the random
reward function in a finite time interval through an event-driven technique.
Furthermore, we calculate the mean of the discounted reward function in an
infinite time interval. Based on this, we give a simple discussion on optimal
criterions for designing the supermarket model with different servers. Also,
we provide some simulation experiments to indicate how the expected queue
length of each server depends on some key parameters of this supermarket model.

The remainder of this paper is organized as follows. In Section 2, we first
describe a supermarket model with $M$ different servers. Then we construct a
routine selection mechanism that depends on the queue lengths, on the service
rates, on the probability of individual preference and so forth. In Section 3,
we set up an $M$-dimensional continuous-time Markov reward process, and
provide a segmented stochastic integral for expressing the random reward
function in a finite time interval through an event-driven technique. In
Section 4, applying the segmented stochastic integral, we compute the mean of
the random reward function in a finite time interval. In Section 5, we compute
the mean of the discounted reward function in an infinite time interval. Based
on this, we provide two optimal criterions for designing the supermarket model
with different servers. In Section 6, we provide some simulation experiments
to indicate how the expected queue length of each server depends on some key
parameters of this supermarket model. Some concluding remarks are given in
Section 7.

\section{Supermarket Model Description}

In this section, we first describe a supermarket model with $M$ different
servers. Then we construct a routine selection mechanism that depends on the
queue lengths, on the service rates, on the probability of individual
preference and so forth.

In the supermarket model, there are $M$ different servers whose waiting rooms
are all infinite. The service times in each server are i.i.d. and are
exponential, and also the service rates of the $M$ different servers are
denoted as $\mu_{1},\mu_{2},\ldots,\mu_{M}$, respectively. The arrivals of
customers are a Poisson process with arrival rate $\lambda$. Because the
servers are different, it is a key to optimize the service ability of this
supermarket model through designing a better routine selection mechanism. In
fact, designing such a better routine selection mechanism will become not only
complicated but also subjective due to the difference of the $M$ servers. The
physical structure of this supermarket model is shown in Figure 1.

In what follows we will provide a detailed description for how to construct
such a better routine selection mechanism. Notice that our method for
constructing the routine selection mechanism is intuitive and heuristic
according to some practical points of view.

\begin{figure}[ptbh]
\centering          \includegraphics[width=9cm]{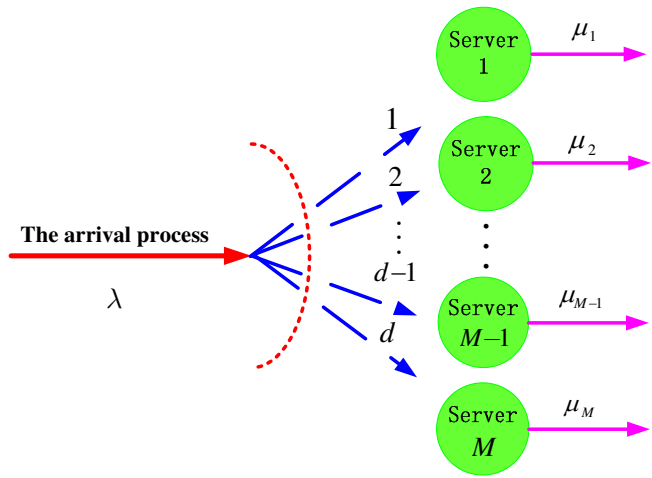}  \caption{A
physical illustration of the supermarket models with different servers}%
\label{figure: fig-1}%
\end{figure}

From Figure 1, it is seen that for the $M$ different servers, each arriving
customer joins a server (or queue) according to a suitable routine selection
mechanism. From a practical point of view, each arriving customer chooses one
server based on at least three crucial factors: (1) Choosing one server with
the largest service rate, (2) choosing one server with the shortest queue
length, and (3) choosing one server with the maximal probability of individual preference.

We write%
\[
x=\left(  x_{1},x_{2},\ldots,x_{M}\right)  ,
\]
which is the vector of the queue lengths in the $M$ servers;%
\[
g=\left(  g_{1},g_{2},\ldots,g_{M}\right)  ,
\]
which is a probability vector of individual preference for choosing one of the
$M$ servers. In general, the individual preference is based on the priori
knownledge, and the present feeling etc.; and%
\[
\mu=\left(  \mu_{1},\mu_{2},\ldots,\mu_{M}\right)  .
\]
It is worth noting that the two vectors $g$ and $\mu$ are always inherent in
the system, but the vector $x$ of queue lengths can change dynamically
according to a customer arrival or a service completion.

Based on the above analysis, let $\Delta_{i}(x)=f(x_{i},\mu_{i},g_{i})$ be a
routine selection function which represents the measurement of choosing the
$i$th server for $i=1,2,\ldots,M$, where $f(x_{i},\mu_{i},g_{i})$ satisfies
three conditions: (1) $f(x_{i},\mu_{i},g_{i})$ is increasing for $x_{i}\geq0$,
(2) $f(x_{i},\mu_{i},g_{i})$ is decreasing for $\mu_{i}>0$, and (3)
$f(x_{i},\mu_{i},g_{i})$ is decreasing for $g_{i}\in(0,1]$.

We assume that if%
\[
\Delta_{i_{0}}(x)=\min_{1\leq i\leq M}\left\{  \Delta_{i}(x)\right\}  ,
\]
then the arriving customer joins the $i_{0}$th server among the $M$ servers.
It indicates that an arriving customer likes the server with the minimal value
in the set of routine selection functions%
\[
\mathbf{\Delta}=\left\{  \Delta_{1}(x),\Delta_{2}(x),\ldots,\Delta
_{M-1}(x),\Delta_{M}(x)\right\}  .
\]

From the routine selection function, now we further describe the routine
selection mechanism as follows:

\textbf{The routine selection mechanism:} Each arriving customer chooses
$d\geq1$ servers independently and uniformly at random from the $M$ servers,
and joins the server with the smallest number in $\mathbf{\Delta}_{d}=\left\{
\Delta_{i_{1}}(x),\Delta_{i_{2}}(x),\ldots,\Delta_{i_{d-1}}(x),\Delta_{i_{d}%
}(x)\right\}  $, where the $d$ selected servers are denoted as Servers
$i_{1},i_{2},\ldots,i_{d}$. If there is a tie, servers with the smallest
number in $\mathbf{\Delta}_{d}$ will be chosen randomly. All customers in any
server will be served in the first come first service (FCFS) manner. We assume
that all the random variables defined for the arrival and service processes
are independent of each other.

In what follows we provide some useful interpretation for each element in the
set $\mathbf{\Delta}=\left\{  \Delta_{1}(x),\Delta_{2}(x),\ldots,\Delta
_{M-1}(x),\Delta_{M}(x)\right\}  $ of routine selection functions.

\textbf{Interpretation one: }$\Delta_{i}(x)=f(x_{i},\mu_{i},g_{i})$\textbf{
has some useful forms}

Note that $f(x_{i},\mu_{i},g_{i})$ needs to satisfy the above three monotone
conditions for each element in one of the three vectors $x$, $\mu$ and $g$,
thus such a function $f:\mathbf{N}^{+}\times(0,+\infty)\times(0,1]\rightarrow
\mathbf{R}^{+}$ can be chosen easily, where $\mathbf{N}^{+}=\left\{
0,1,2,\ldots\right\}  $ and $\mathbf{R}^{+}=[0,+\infty)$. To that end, we give
some examples to indicate how to construct such a function $f(x_{i},\mu
_{i},g_{i})$ as follows:

(1) A tandem-type decision-making method

For the three decision variables $x_{i}$, $\mu_{i}$ and $g_{i}$, we set up a
tandem-type decision-making structure as $x_{i}\cdot\frac{1}{\mu_{i}}%
\cdot\frac{1}{g_{i}}$, thus it is seen from a normalization that the routine
selection function is given by%
\[
\Delta_{i}(x)=\frac{1+\frac{x_{i}}{\mu_{i}g_{i}}}{\underset{j=1}{\overset
{M}{\sum}}\left[  1+\frac{x_{j}}{\mu_{j}g_{j}}\right]  },\text{ \ }%
i=1,2,\ldots,M.
\]

(2) A weighted-type decision-making method

For the three decision variables $x_{i}$, $\mu_{i}$ and $g_{i}$, we take a
weighted-type decision-making structure as $\beta_{1}x_{i}+\beta_{2}%
\frac{1}{\mu_{i}}+\beta_{3}\frac{1}{g_{i}}$, where the weighted coefficients
satisfy that $\beta_{k}\geq0$ and $\beta_{1}+\beta_{2}+\beta_{3}=1$, thus the
routine selection function is given by%
\[
\Delta_{i}(x)=\frac{1+\beta_{1}x_{i}+\beta_{2}\frac{1}{\mu_{i}}+\beta
_{3}\frac{1}{g_{i}}}{\underset{j=1}{\overset{M}{\sum}}\left[  1+\beta_{1}%
x_{j}+\beta_{2}\frac{1}{\mu_{j}}+\beta_{3}\frac{1}{g_{j}}\right]  },\text{
\ }i=1,2,\ldots,M.
\]

\textbf{Interpretation two: There exist multiple minimal elements in
}$\mathbf{\Delta}_{d}$

For $\mathbf{\Delta}_{d}=\left\{  \Delta_{i_{1}}(x),\Delta_{i_{2}}%
(x),\ldots,\Delta_{i_{d-1}}(x),\Delta_{i_{d}}(x)\right\}  $, set%
\[
\Re_{\min}\left(  d\right)  =\left\{  i_{0}:\Delta_{i_{0}}(x)=\min_{1\leq
k\leq d}\left\{  \Delta_{i_{k}}(x)\right\}  \right\}  .
\]
Then we have two cases: (1) $\Re_{\min}\left(  d\right)  $ contains only one
element, and (2) $\Re_{\min}\left(  d\right)  $ contains multiple elements.
For the former, the routine selection of the arriving customer is simple for
choosing Server $i_{0}$; while for the latter, the routine selection of the
arriving customer has a little complicated, for example, a simple mode is
taken as that if there is a tie, servers with the smallest number in
$\Re_{\min}\left(  d\right)  $ will be chosen randomly, e.g., see Vvedenskaya
et al. \cite{Vve:1996} and Mitzenmacher \cite{Mit:1996}.

To use more information in the set $\Re_{\min}\left(  d\right)  $, we may set
up some new routine selection ways. If there is a tie (that is, $\Re_{\min
}\left(  d\right)  $ contains multiple elements), then servers with the
smallest number in $\Re_{\min}\left(  d\right)  $ may be chosen by means of
other ways, for example, either

(1) for all the different elements in $\Re_{\min}\left(  d\right)  $, the
arriving customer joins the server with the biggest service rate;

(2) for all the different elements in $\Re_{\min}\left(  d\right)  $, the
arriving customer joins the server with the shortest queue length;

(3) for all the different elements in $\Re_{\min}\left(  d\right)  $, the
arriving customer joins the server with the maximal probability of individual
preference; or

(4) some hybrid combination from the above (1), (2) and (3).

In this paper, we will not discuss the above four cases, which are interesting
and will be studied in our future work.

\textbf{Interpretation three: Useful relations between our above model and the
ordinary supermarket model}

On the one hand, when $\mu_{1}=\mu_{2}=\cdots=\mu_{M}=\mu$ and $g_{1}=g_{2}=\cdots
=g_{M}=\frac{1}{M}$, it is seen that%
\[
\Delta_{i}(x)=f(x_{i},\mu_{i},g_{i})=f\left(  x_{i},\mu,\frac{1}{M}\right)  ,
\]
which shows that the routine selection of the arriving customer only depends
on the vector $x=\left(  x_{1},x_{2},\ldots,x_{M}\right)  $, hence the
arriving customer joins the server with the shortest queue length, e.g., see
Vvedenskaya et al. \cite{Vve:1996}. On the other hand, we remark that the probability vector
$g=\left(  g_{1},g_{2},\ldots,g_{M}\right)  $ of individual preference can
give rise to the study of modern supermarket business or network economy.

In the supermarket model with different servers, data collection and analysis is
also a key task. Therefore, it is interesting that the routine selection mechanism
can be designed from a data-based practical point of view. This will motivate
statistical analysis of supermarket models with different servers from many real areas.

\section{A Markov Reward Process}

In this section, we set up an $M$-dimensional continuous-time Markov reward
process, and provide a segmented stochastic integral for expressing the random
reward function in a finite time interval through an event-driven technique.

In order to set up a continuous-time Markov reward process, we need to discuss
the arrival and service processes, both of which lead to the state jumps of
this Markov reward process. At the same time, we choose a suitable reward
value at each state in this supermarket model.

\textbf{(1) Analysis of the arrival processes}

In this supermarket model, the arrival process of customers is a Poisson
process with arrival rate $\lambda$. Each arriving customer chooses $d$
servers independently and uniformly at random from the $M$ servers, and joins
one server with the smallest number in the set $\mathbf{\Delta}_{d}=\left\{
\Delta_{i_{1}}(x),\Delta_{i_{2}}(x),\ldots,\Delta_{i_{d-1}}(x),\Delta_{i_{d}%
}(x)\right\}  $. If there is a tie, servers with the smallest number in the
set $\Delta_{d}$ will be chosen randomly.

In order to express the routine selection mechanism of each arriving customer,
we need to introduce an \textit{ascending} function $\sigma:\left[
0,1\right]  ^{M}\rightarrow\left[  0,1\right]  ^{M}$ as follows:%
\[
\sigma\left(  \mathbf{\Delta}^{\left(  x\right)  }\right)  =\left(
\Delta_{k_{1}}(x),\Delta_{k_{2}}(x),\ldots,\Delta_{k_{M}}(x)\right)
\]
for $\mathbf{\Delta}^{\left(  x\right)  }=\left(  \Delta_{1}(x),\Delta
_{2}(x),\ldots,\Delta_{M}(x)\right)  $, where%
\begin{equation}
0\leq\Delta_{k_{1}}(x)\leq\Delta_{k_{2}}(x)\leq\ldots\leq\Delta_{k_{M}}%
(x)\leq1. \label{Equ3-2}%
\end{equation}

For the ascending function $\sigma\left(  \mathbf{\Delta}^{\left(  x\right)
}\right)  $, it is necessary to explain the order numbers $k_{i}$ for $1\leq
i\leq M$. Note that $k_{i}$ denotes the $k_{i}$th element of the original
order number vector $\mathbf{\Delta}^{\left(  x\right)  }$. For example, if
$\mathbf{\Delta}^{\left(  x\right)  }=\left(  1/3,1/2,1/6\right)  $, then
$\sigma\left(  \mathbf{\Delta}^{\left(  x\right)  }\right)  =\left(
1/6,1/3,1/2\right)  $. It is obvious that $\Delta_{k_{1}}(x)=1/6$ and
$k_{1}=3$; $\Delta_{k_{2}}(x)=1/3$ and $k_{2}=1$; and $\Delta_{k_{3}}(x)=1/2$
and $k_{3}=2$. In general, for these order numbers before and after sorting,
we provide their corresponding relation in Figure 2.

\begin{figure}[ptbh]
\centering           \includegraphics[width=9cm]{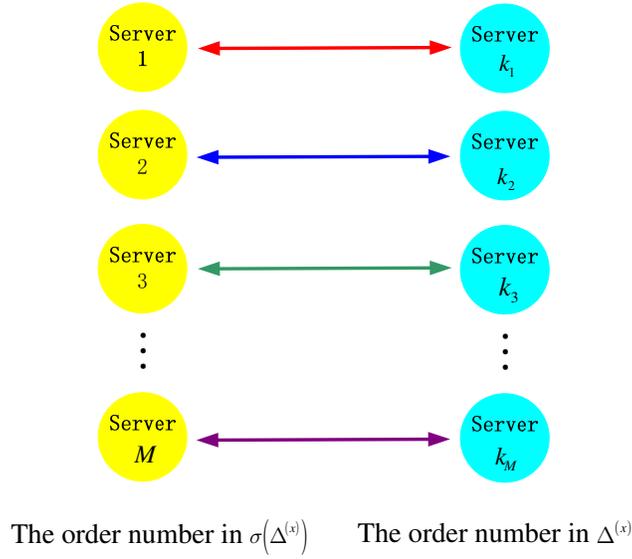}  \caption{The
order relation before and after sorting the $M$ servers}%
\label{figure: fig-2}%
\end{figure}

Based on the ascending function with the sorting process, it is a key how to
describe the arrivals of customers at each server in this supermarket model.
It is worthwhile to note that Janssen \cite{Jan:2011} gave an effective method
for analyzing the ascending function as well as the arrival processes at the
$M$ different servers. Here, we provide a detailed description for the
Janssen's method as follows.

For a sorted vector $x$ with $0\leq x_{1}\leq x_{2}\leq\cdots\leq x_{M}$, it
follows from (3.6) and (3.7) in Janssen \cite{Jan:2011} that the probability
that the arriving customer first randomly selects $d$ servers from $M$
servers, and then enters the $i$th server (that is, the $i$th shortest queue
is also in the $d$ selected servers) is given by%
\begin{equation}
k(M,i,d)=\left\{
\begin{array}
[c]{ll}%
d\frac{(M-i)!(M-d)!}{(M-i-d+1)!(M)!}, & \text{ \ }1\leq i\leq M-d+1,\\
0, & \text{ \ }M-d+2\leq i\leq M,
\end{array}
\right.  \label{Equ3-3}%
\end{equation}
and specifically, we may randomly give a sort for these servers whose queue
lengths are equal. At the same time, Lemma 3.2.1 in Janssen \cite{Jan:2011}
proved that for $1\leq d\leq M$,%
\begin{equation}
\underset{i=1}{\overset{M}{\sum}}k(M,i,d)=\underset{i=1}{\overset{M-d+1}{\sum
}}k(M,i,d)=1. \label{Equ3-4}%
\end{equation}

Now, we explain the probability $k(M,i,d)$ for sorted vector $x$ with $0\leq
x_{1}\leq x_{2}\leq\cdots\leq x_{M}$.

As seen from Figure 3, notice that the arriving customer first randomly
selects $d$ servers from the $M$ servers, and enters one server with the
shortest queue length among the $d$ selected servers (if there is a tie, then
servers with the shortest queue length will be chosen randomly), thus the
routine selection mechanism is converted to the probability $k(M,i,d)$ of
entering the $i$th server for $1\leq i\leq M$. Therefore, $\lambda k(M,i,d)$
is the arrival rate that the customers arrive at the server with the $i$th
shortest queue length among the $M$ servers.

\begin{figure}[ptbh]
\centering             \includegraphics[width=10cm]{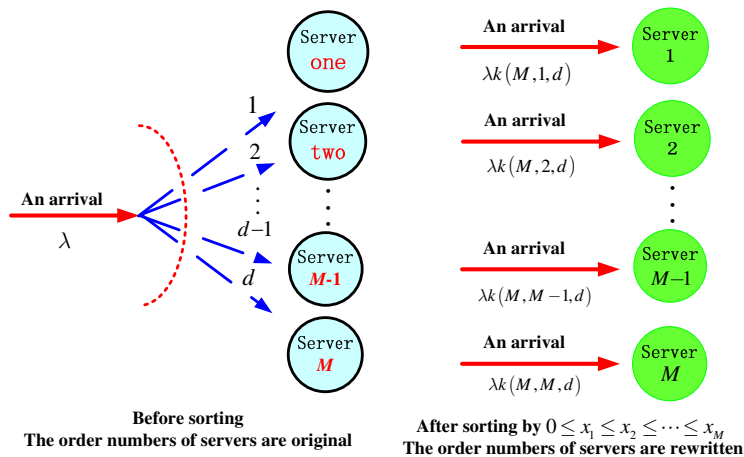}  \caption{Some
interpretation on the probability $k(M,i,d)$}%
\label{figure: fig-3}%
\end{figure}

For the ascending function $\sigma\left(  \mathbf{\Delta}^{\left(  x\right)
}\right)  $ which is similar to the sorted vector $x$ with $0\leq x_{1}\leq
x_{2}\leq\cdots\leq x_{M}$, it is easy to see that the Janssen's method still
work. Thus, for the $i$th element in $\sigma\left(  \mathbf{\Delta}^{\left(
x\right)  }\right)  $ (that is, the $k_{i}$th element in $\mathbf{\Delta
}^{\left(  x\right)  }$, this corresponds to the $k_{i}$th server in this
supermarket model), using (\ref{Equ3-4}) we obtain%
\[
k(M,k_{i},d)=\left\{
\begin{array}
[c]{ll}%
d\frac{(M-k_{i})!(M-d)!}{(M-k_{i}-d+1)!(M)!}, & \text{ \ }1\leq k_{i}\leq
M-d+1,\\
0, & \text{ \ }M-d+2\leq k_{i}\leq M.
\end{array}
\right.
\]
Obviously, we also have%
\[
\underset{k_{i}=1}{\overset{M}{\sum}}k(M,k_{i},d)=\underset{k_{i}=1}%
{\overset{M-d+1}{\sum}}k(M,k_{i},d)=1.
\]

According to the probability $k(M,k_{i},d)$, it is clear that the arrivals of
customers at the $k_{i}$th server is a Poisson process with arrival rate
$\lambda k(M,k_{i},d)$ for $i=1,2,\ldots,M$. Hence, the Poisson arrival rate
at the $k_{i}$th server is given by%
\begin{equation}
\lambda k(M,k_{i},d)=\left\{
\begin{array}
[c]{ll}%
\lambda d\frac{(M-k_{i})!(M-d)!}{(M-k_{i}-d+1)!(M)!}, & \text{ \ }1\leq
k_{i}\leq M-d+1,\\
0, & \text{ \ }M-d+2\leq k_{i}\leq M.
\end{array}
\right.  \label{Eq3-5}%
\end{equation}

\textbf{(2) Analysis of the service processes }

Analysis of the service processes is simpler than that of the above arrival
processes in this supermarket model. Let $\mathbf{1}_{\left\{  x_{i}%
>0\right\}  }$ be an indicator function of the event: $\left\{  x_{i}%
>0\right\}  $, that is,%
\[
\mathbf{1}_{\left\{  x_{i}>0\right\}  }=\left\{
\begin{array}
[c]{cc}%
1, & \text{ \ }x_{i}>0,\\
0, & \text{ \ }x_{i}=0.
\end{array}
\right.
\]
The service rate of the $i$th server may be written as $\mu_{i}\mathbf{1}%
_{\left\{  x_{i}>0\right\}  }$, because the server is idle when there is no
customer (i.e., $x_{i}=0$) in this server.

\textbf{(3) Choosing a suitable reward value at each state}

Note that $\Delta_{k_{M}}(x)\geq\Delta_{k_{1}}(x)$, it is obvious that if the
value $\left[  \Delta_{k_{M}}(x)-\Delta_{k_{1}}(x)\right]  /\Delta_{k_{M}}(x)$
is bigger, then the customers in the $M$ servers are not distributed well. On
the contrary, if the value $\left[  \Delta_{k_{M}}(x)-\Delta_{k_{1}%
}(x)\right]  /\Delta_{k_{M}}(x)$ is smaller, then the customers in the $M$
servers are load balanced very well. Thus, our purpose of designing and
optimizing this supermarket model is to make the value $\left[  \Delta_{k_{M}%
}(x)-\Delta_{k_{1}}(x)\right]  /\Delta_{k_{M}}(x)$ as small as possible. At
the same time, it is easy to see that%
\[
\underset{d;\lambda;\mu_{k},1\leq k\leq M}{\min}\left\{  \Delta_{k_{M}%
}(x)\right\}  -\underset{d;\lambda;\mu_{k},1\leq k\leq M}{\max}\left\{
\Delta_{k_{1}}(x)\right\}  \leq\Delta_{k_{M}}(x)-\Delta_{k_{1}}(x).
\]
Based on the above analysis, we may choose two different reward values at
state $x$ as follows:%
\begin{equation}
r_{\min}(x):=\Delta_{k_{1}}(x), \label{Eq3-6}%
\end{equation}
and%
\begin{equation}
r_{\max}(x):=\Delta_{k_{M}}(x). \label{Eq3-7}%
\end{equation}
Notice that we use the two reward values: $r_{\min}(x)$ and $r_{\max}(x)$, to
be able to provide a better observation on performance of this supermarket
model, which will be studied in Subsection 5.2.

In the remainder of this section, we introduce a useful continuous-time Markov
process, which will be used to give performance computation and performance
simulation in the supermarket model with different servers.

Let $X_{k}\left(  t\right)  $ be the number of customers in the $k$th server
of this supermarket model at time $t\geq0$, and%
\[
\mathbf{X}\left(  t\right)  =\left(  X_{1}\left(  t\right)  ,X_{2}\left(
t\right)  ,\ldots,X_{M}\left(  t\right)  \right)  .
\]
Obviously, $\left\{  \mathbf{X}\left(  t\right)  :t\geq0\right\}  $ is an
$M$-dimensional continuous-time Markov process on the space state
$\Omega=\left\{  x=\left(  x_{1},x_{2},\ldots,x_{M}\right)  :x_{k}\geq0,1\leq
k\leq M\right\}  $.

Let $r\left(  x\right)  $ be a real function for $x\in\Omega$, and $r\left(
x\right)  $ denote a reward value of this Markov process $\left\{
\mathbf{X}\left(  t\right)  :t\geq0\right\}  $ at state $x$. Based on this, we
define a random reward function as%
\begin{equation}
\Phi\left(  t\right)  =\int_{0}^{t}r\left(  \mathbf{X}\left(  \xi\right)
\right)  \text{d}\xi, \label{Eq3-8}%
\end{equation}
which is a stochastic integral, e.g., see Chapter 10 in Li \cite{Li:2010} for
more details.

In what follows we propose an event-driven technique to deal with the random
reward function $\Phi\left(  t\right)  $. To this end, we denote by $\eta
_{1},\eta_{2},\eta_{3},\ldots,\eta_{n}$ the $n$ successive state jump points
of the Markov process $\left\{  \mathbf{X}\left(  t\right)  :t\geq0\right\}  $
in the finite time interval $\left[  0,t\right]  $, it is clear that%
\begin{equation}
0<\eta_{1}<\eta_{2}<\cdots<\eta_{n}<t<\eta_{n+1}. \label{Eq3-9}%
\end{equation}
Note that $\eta_{k}=\eta_{k}^{-}$, and $\eta_{k}$ is a state jump time of
Markov process $\left\{  \mathbf{X}\left(  t\right)  :t\geq0\right\}  $, thus
it is helpful for understanding the stochastic integral $\int_{0}^{t}r\left(
\mathbf{X}\left(  \xi\right)  \right)  $d$\xi$ under an interval decomposition
as follows:
\[
\left[  0,t\right]  =[0,\eta_{1}^{-})\cup\lbrack\eta_{1},\eta_{2}^{-}%
)\cup\lbrack\eta_{2},\eta_{3}^{-})\cdots\cup\lbrack\eta_{n-1},\eta_{n}%
^{-})\cup\lbrack\eta_{n},t],
\]
it follows from (\ref{Eq3-8}) and (\ref{Eq3-9}) that%
\begin{equation}
\Phi\left(  t\right)  =\int_{0}^{\eta_{1}^{-}}r\left(  \mathbf{X}\left(
\xi\right)  \right)  \text{d}\xi+\sum_{j=1}^{n-1}\int_{\eta_{j}}^{\eta
_{j+1}^{-}}r\left(  \mathbf{X}\left(  \xi\right)  \right)  \text{d}\xi
+\int_{\eta_{n}}^{t}r\left(  \mathbf{X}\left(  \xi\right)  \right)
\text{d}\xi, \label{Eq3-10}%
\end{equation}
which is a segmented stochastic integral for expressing the random reward
function $\Phi\left(  t\right)  $. Note that this segmented stochastic
integrals will be useful in our later study.

\section{Computation of the Expected Reward Function}

In this section, we use an event-driven technique to compute the mean of the
random reward function in a finite time interval, where our computation is
based on the above segmented stochastic integral, which is expressed through
the successive state jump points generated by either customer arrivals or
service completions.

From (a) in Figure 4, let $\left\{  \mathcal{N}\left(  t\right)
:t\geq0\right\}  $ be a Poisson process with parameter $\omega=\lambda+\mu
_{1}+\mu_{2}+\cdots+\mu_{M}$. Then for $k\geq0$%
\[
\mathbf{p}_{k}\left(  t\right)  =P\left\{  \mathcal{N}\left(  t\right)
=k\right\}  =e^{-\omega t}\frac{\left(  \omega t\right)  ^{k}}{k!}.
\]

\begin{figure}[ptbh]
\centering             \includegraphics[width=10cm]{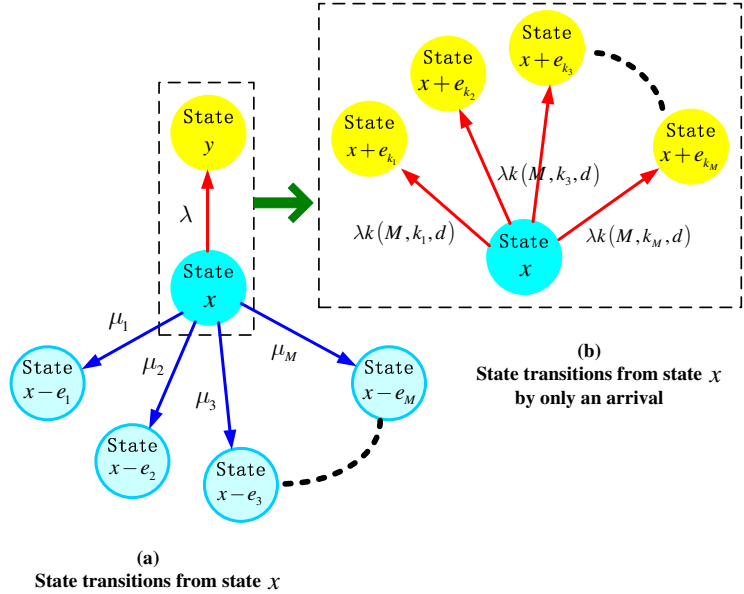}  \caption{State
transitions and associated rates at State $x$}%
\label{figure: fig-4}%
\end{figure}

We assume that the random sequence $\left\{  Y_{k}:k\geq1\right\}  $ is i.i.d.
and is exponential with mean $1/\omega$. Let $\eta_{n}=\sum_{k=1}^{n}Y_{k}$.
Then $\mathcal{N}\left(  t\right)  =\sup\left\{  n:\eta_{n}\leq t\right\}  $,
and $0<\eta_{1}<\eta_{2}<\cdots<\eta_{n}<t<\eta_{n+1}$. From Section 2.3 in
Ross \cite{Ross:1983}, it is easy to see that%
\[
P\left\{  \eta_{1}\leq s\text{ }|\text{ }\mathcal{N}\left(  t\right)
=1\right\}  =P\left\{  Y_{1}\leq s\text{ }|\text{ }\mathcal{N}\left(
t\right)  =1\right\}  =\frac{s}{t}.
\]
Let the $n$-dimensional probability distribution be%
\[
F\left(  s_{1},s_{2},\ldots,s_{n}\right)  =P\left\{  \eta_{1}\leq s_{1}%
,\eta_{2}\leq s_{2},\ldots,\eta_{n}\leq s_{n}\right\}
\]
and the $n$-dimensional probability density function%
\[
f\left(  s_{1},s_{2},\ldots,s_{n}\right)  =\frac{\partial^{n}}{\partial
s_{1}\partial s_{2}\cdots\partial s_{n}}F\left(  s_{1},s_{2},\ldots
,s_{n}\right)  .
\]
Then it follows from Theorem 2.3.1 in Ross \cite{Ross:1983} that%
\[
f\left(  s_{1},s_{2},\ldots,s_{n}\right)  =\frac{n!}{t^{n}},\text{ \ }%
0<s_{1}<s_{2}<\cdots<s_{n}<t.
\]
At the same time, Theorem 2.3.1 in Ross \cite{Ross:1983} demonstrates that
given that $\mathcal{N}\left(  t\right)  =n$, the $n$ arrival times $\eta
_{1},\eta_{2},\ldots,\eta_{n}$ have the same distribution as the order
statistics corresponding to $n$ independent random variables uniformly
distributed on the interval $\left(  0,t\right)  $. Thus, using the condition:
$0<\eta_{1}<\eta_{2}<\cdots<\eta_{n}<t$, we obtain%
\begin{equation}
E\left[  \eta_{1}\right]  =E\left[  \eta_{2}-\eta_{1}\right]  =\cdots=E\left[
\eta_{n}-\eta_{n-1}\right]  =E\left[  t-\eta_{n}\right]  =\frac{t}{n+1}.
\label{4.0}%
\end{equation}

It is seen from (a) and (b) in Figure 4 that for $k\geq1$, the Markov process
$\left\{  \mathbf{X}\left(  t\right)  :t\geq0\right\}  $ transits to State
$\mathbf{X}\left(  \eta_{k}\right)  $ from State $\mathbf{X}\left(  \eta
_{k}^{-}\right)  $ (i.e., a state jump), where State $\mathbf{X}\left(
\eta_{k}\right)  $ may be either State $\mathbf{X}\left(  \eta_{k}^{-}\right)
-e_{j}$ due to a service completion by Server $j$ for $1\leq j\leq M$, or
State $\mathbf{X}\left(  \eta_{k}^{-}\right)  +e_{k_{i}}$ due to a customer
arrival at Server $k_{i}$ with the routine selection mechanism for $1\leq
i\leq M$. Note that $\mathbf{X}\left(  \eta_{1}^{-}\right)  =\mathbf{X}\left(
0\right)  =x$ and $\mathbf{X}\left(  \eta_{k}^{-}\right)  =\mathbf{X}\left(
\eta_{k-1}\right)  $ for $2\leq k\leq n$, thus we have%
\begin{align*}
\mathbf{X}\left(  \eta_{k}\right)   &  \in\left\{  \mathbf{X}\left(  \eta
_{k}^{-}\right)  -e_{j}:1\leq j\leq M\right\}  \cup\left\{  \mathbf{X}\left(
\eta_{k}^{-}\right)  +e_{k_{i}}:1\leq i\leq M\right\} \\
&  =\left\{  \mathbf{X}\left(  \eta_{k-1}\right)  -e_{j}:1\leq j\leq
M\right\}  \cup\left\{  \mathbf{X}\left(  \eta_{k-1}\right)  +e_{k_{i}}:1\leq
i\leq M\right\}  .
\end{align*}

Let $A_{n}$ be the $n$th inter-arrival time of the Poisson process with
arrival rate $\lambda$, and $S_{n}^{\left(  k\right)  }$ the exponential
service time with service rate $\mu_{k}$ of the $n$th customer in Server $k$.
Then $\left\{  A_{n}\right\}  $ and $\left\{  S_{n}^{\left(  k\right)
}\right\}  $ are all i.i.d for $1\leq k\leq M$. In this case, we write that
$A=A_{1}$ and $S^{\left(  k\right)  }=S_{1}^{\left(  k\right)  }$ for $1\leq
k\leq M$. Based on these random variables $A$ and $S^{\left(  k\right)  }$ for
$1\leq k\leq M$, we can express the random events of the Markov process
$\left\{  \mathbf{X}\left(  t\right)  :t\geq0\right\}  $ at time $\eta_{k}$ as follows:

(1) An arrival at time $\eta_{k}$

In this case, we need the sufficient condition%
\[
A<\min_{1\leq k\leq M}\left\{  S^{\left(  k\right)  }\right\}  .
\]
It is easy to compute that%
\[
a=P\left\{  A<\min_{1\leq k\leq M}\left\{  S^{\left(  k\right)  }\right\}
\right\}  =\frac{\lambda}{\lambda+\mu_{1}+\mu_{2}+\cdots+\mu_{M}}.
\]

(2) A service completion in Server $j$ for $1\leq j\leq M$

In this case, we need the sufficient condition%
\[
S^{\left(  j\right)  }<\min\left\{  A,\min_{\substack{k\neq j\\1\leq k\leq
M}}\left\{  S^{\left(  k\right)  }\right\}  \right\}  .
\]
We can that%
\[
b^{\left(  j\right)  }=P\left\{  S^{\left(  j\right)  }<\min\left\{
A,\min_{\substack{k\neq j\\1\leq k\leq M}}\left\{  S^{\left(  k\right)
}\right\}  \right\}  \right\}  =\frac{\mu_{j}}{\lambda+\mu_{1}+\mu_{2}%
+\cdots+\mu_{M}}.
\]

Now, we compute the conditional mean $E_{x}\left[  \Phi\left(  t\right)
\right]  $, where $E_{x}\left[  \bullet\right]  =E\left[  \bullet\text{
}|\text{ }\mathbf{X}\left(  0\right)  =x\right]  $.

We have%
\begin{align}
E\left[  \Phi\left(  t\right)  \text{ }|\text{ }\mathbf{X}\left(  0\right)
=x\right]  =  &  \sum_{n=0}^{\infty}P\left\{  \mathcal{N}\left(  t\right)
=n\right\}  E\left[  \Phi\left(  t\right)  \text{ }|\text{ }\mathbf{X}\left(
0\right)  =x,\mathcal{N}\left(  t\right)  =n\right] \nonumber\\
=  &  P\left\{  \mathcal{N}\left(  t\right)  =0\right\}  E\left[  \Phi\left(
t\right)  \text{ }|\text{ }\mathbf{X}\left(  0\right)  =x,\mathcal{N}\left(
t\right)  =0\right] \nonumber\\
&  +\sum_{n=1}^{\infty}P\left\{  \mathcal{N}\left(  t\right)  =n\right\}
E\left[  \Phi\left(  t\right)  \text{ }|\text{ }\mathbf{X}\left(  0\right)
=x,\mathcal{N}\left(  t\right)  =n\right]  , \label{4.1}%
\end{align}

Since $\mathcal{N}\left(  t\right)  =0$, it is clear that $\eta_{1}>t$, this
gives
\begin{equation}
E\left[  \Phi\left(  t\right)  \text{ }|\text{ }\mathbf{X}\left(  0\right)
=x,\mathcal{N}\left(  t\right)  =0\right]  =E\left[  \int_{0}^{t}r\left(
\mathbf{X}\left(  \xi\right)  \right)  \text{d}\xi\text{ }|\text{ }%
\mathbf{X}\left(  0\right)  =x,\eta_{1}>t\right]  =r\left(  x\right)  t.
\label{4.2}%
\end{equation}

For $n\geq1$, notice that the event $\left\{  \mathcal{N}\left(  t\right)
=n\right\}  $ is the same as the event $\left\{  0<\eta_{1}<\eta_{2}%
<\cdots\right.  $ $\left.  <\eta_{n}<t<\eta_{n+1}\right\}  $, thus we obtain%
\begin{align}
&  E\left[  \Phi\left(  t\right)  \text{ }|\text{ }\mathbf{X}\left(  0\right)
=x,\mathcal{N}\left(  t\right)  =n\right] \nonumber\\
&  =E\left[  \int_{0}^{t}r\left(  \mathbf{X}\left(  \xi\right)  \right)
\text{d}\xi\text{ }|\text{ }\mathbf{X}\left(  0\right)  =x,0<\eta_{1}<\eta
_{2}<\cdots<\eta_{n}<t<\eta_{n+1}\right] \nonumber\\
&  =E\left[  \int_{0}^{\eta_{1}^{-}}r\left(  \mathbf{X}\left(  \xi\right)
\right)  \text{d}\xi\text{ }|\text{ }\mathbf{X}\left(  0\right)  =x,0<\eta
_{1}<\eta_{2}<\cdots<\eta_{n}<t<\eta_{n+1}\right] \nonumber\\
&  +\sum_{k=1}^{n-1}E\left[  \int_{\eta_{k}}^{\eta_{k+1}^{-}}r\left(
\mathbf{X}\left(  \xi\right)  \right)  \text{d}\xi\text{ }|\text{ }%
\mathbf{X}\left(  0\right)  =x,0<\eta_{1}<\eta_{2}<\cdots<\eta_{n}%
<t<\eta_{n+1}\right] \nonumber\\
&  +E\left[  \int_{\eta_{n}}^{t}r\left(  \mathbf{X}\left(  \xi\right)
\right)  \text{d}\xi\text{ }|\text{ }\mathbf{X}\left(  0\right)  =x,0<\eta
_{1}<\eta_{2}<\cdots<\eta_{n}<t<\eta_{n+1}\right]  \label{4.3}%
\end{align}

To compute (\ref{4.3}), we may observe some useful relations as follows:

(1) If $\xi\in\lbrack0,\eta_{1}^{-})$ and $\mathbf{X}\left(  0\right)  =x$,
then $\mathbf{X}\left(  \xi\right)  =x$ for $\xi\in\lbrack0,\eta_{1}^{-})$.

(2) For $1\leq j\leq n-1$, if $\xi\in\lbrack\eta_{j},\eta_{j+1}^{-})$ and
$\mathbf{X}\left(  \eta_{j}\right)  =y$, then $\mathbf{X}\left(  \xi\right)
=y$ for $\xi\in\lbrack\eta_{j},\eta_{j+1}^{-})$.

(3) If $\xi\in\lbrack\eta_{n},t]$ and $\mathbf{X}\left(  \eta_{n}\right)  =z$,
then $\mathbf{X}\left(  \xi\right)  =z$ for $\xi\in\lbrack\eta_{n},t]$.

Based on the above useful relations, together with (\ref{4.0}), we obtain%
\begin{align*}
&  E\left[  \int_{0}^{\eta_{1}^{-}}r\left(  \mathbf{X}\left(  \xi\right)
\right)  \text{d}\xi\text{ }|\text{ }\mathbf{X}\left(  0\right)  =x,0<\eta
_{1}<\eta_{2}<\cdots<\eta_{n}<t<\eta_{n+1}\right] \\
&  =r\left(  x\right)  E\left[  \eta_{1}^{-}\right]  =r\left(  x\right)
E\left[  \eta_{1}\right]  =r\left(  x\right)  \frac{t}{n+1},
\end{align*}
for $1\leq k\leq n-1$%
\begin{align*}
&  E\left[  \int_{\eta_{k}}^{\eta_{k+1}^{-}}r\left(  \mathbf{X}\left(
\xi\right)  \right)  \text{d}\xi\text{ }|\text{ }\mathbf{X}\left(  0\right)
=x,0<\eta_{1}<\eta_{2}<\cdots<\eta_{n}<t<\eta_{n+1}\right] \\
&  =E\left[  r\left(  \mathbf{X}\left(  \eta_{k}\right)  \right)  \text{
}|\text{ }\mathbf{X}\left(  0\right)  =x,0<\eta_{1}<\eta_{2}<\cdots<\eta
_{n}<t<\eta_{n+1}\right]  \cdot E\left[  \eta_{k+1}^{-}-\eta_{k}\right] \\
&  =E\left[  r\left(  \mathbf{X}\left(  \eta_{k}\right)  \right)  \text{
}|\text{ }\mathbf{X}\left(  0\right)  =x,0<\eta_{1}<\eta_{2}<\cdots<\eta
_{n}<t<\eta_{n+1}\right]  \cdot E\left[  \eta_{k+1}-\eta_{k}\right] \\
&  =\frac{t}{n+1}E\left[  r\left(  \mathbf{X}\left(  \eta_{k}\right)  \right)
\text{ }|\text{ }\mathbf{X}\left(  0\right)  =x,0<\eta_{1}<\eta_{2}%
<\cdots<\eta_{n}<t<\eta_{n+1}\right]
\end{align*}
and%
\begin{align*}
&  E\left[  \int_{\eta_{n}}^{t}r\left(  \mathbf{X}\left(  \xi\right)  \right)
\text{d}\xi\text{ }|\text{ }\mathbf{X}\left(  0\right)  =x,0<\eta_{1}<\eta
_{2}<\cdots<\eta_{n}<t<\eta_{n+1}\right] \\
&  =E\left[  r\left(  \mathbf{X}\left(  \eta_{n}\right)  \right)  \text{
}|\text{ }\mathbf{X}\left(  0\right)  =x,0<\eta_{1}<\eta_{2}<\cdots<\eta
_{n}<t<\eta_{n+1}\right]  \cdot E\left[  t-\eta_{n}\right] \\
&  =\frac{t}{n+1}E\left[  r\left(  \mathbf{X}\left(  \eta_{n}\right)  \right)
\text{ }|\text{ }\mathbf{X}\left(  0\right)  =x,0<\eta_{1}<\eta_{2}%
<\cdots<\eta_{n}<t<\eta_{n+1}\right]  .
\end{align*}
We write that for $1\leq k\leq n$%
\[
\Re_{k}=E\left[  r\left(  \mathbf{X}\left(  \eta_{k}\right)  \right)  \text{
}|\text{ }\mathbf{X}\left(  0\right)  =x,0<\eta_{1}<\eta_{2}<\cdots<\eta
_{n}<t<\eta_{n+1}\right]  .
\]
It follows from (\ref{4.1}), (\ref{4.2}) and (\ref{4.3}) that%
\begin{equation}
E\left[  \Phi\left(  t\right)  \text{ }|\text{ }\mathbf{X}\left(  0\right)
=x\right]  =r\left(  x\right)  te^{-\omega t}+\sum_{n=1}^{\infty}e^{-\omega
t}\frac{\left(  \omega t\right)  ^{n}}{n!}\cdot\frac{t}{n+1}\left[  r\left(
x\right)  +\sum_{k=1}^{n}\Re_{k}\right]  . \label{Eq4.1}%
\end{equation}
Clearly, it is a key to compute the functions: $\Re_{k}$ for $1\leq k\leq n$.

Now, we use (\ref{Eq4.1}) to compute the conditional mean $E\left[
\Phi\left(  t\right)  \text{ }|\text{ }\mathbf{X}\left(  0\right)  =x\right]
$ of the random reward function $\Phi\left(  t\right)  $\ through an
event-driven technique. To this end, our computation is decomposed in the
following three steps:

\textbf{Step one: Compute }$\Re_{1}=E\left[  r\left(  \mathbf{X}\left(
\eta_{1}\right)  \right)  \text{ }|\text{ }\mathbf{X}\left(  0\right)
=x,0<\eta_{1}<\eta_{2}<\cdots<\eta_{n}<t<\eta_{n+1}\right]  $

It is seen from (a) and (b) in Figure 4 that the Markov process $\left\{
\mathbf{X}\left(  t\right)  :t\geq0\right\}  $ transits to a state
$\mathbf{X}\left(  \eta_{1}\right)  $ from the initial state $x$, where the
state $\mathbf{X}\left(  \eta_{1}\right)  $ may be either State $x-e_{j}$ due
to a service completion by Server $j$ for $1\leq j\leq M$, or State
$x+e_{k_{i}}$ due to a customer arrival at Server $k_{i}$ for $1\leq i\leq M$.
Using the routine selection mechanism, we have%
\[
\mathbf{X}\left(  \eta_{1}\right)  \in\left\{  x-e_{j}:1\leq j\leq M\right\}
\cup\left\{  x+e_{k_{i}}:1\leq i\leq M\right\}  .
\]

From (a) and (b) in Figure 4, it is seen that the computation of\textbf{ }%
$\Re_{1}$ is decomposed into two parts: One by an arrival, and another by a
service completion. Thus we obtain
\begin{equation}
\Re_{1}=\sum_{i=1}^{M}r\left(  x+e_{k_{i}}\right)  \cdot a\cdot k\left(
M,k_{i},d\right)  +\sum_{j=1}^{M}r\left(  x-e_{j}\right)  \cdot b^{\left(
j\right)  }\mathbf{1}_{\left\{  x_{j}>0\right\}  }, \label{Eq4.2}%
\end{equation}
where $a\cdot k\left(  M,k_{i},d\right)  $ is the probability that an arriving
customer joins Server $k_{i}$, and $b^{\left(  j\right)  }\mathbf{1}_{\left\{
x_{j}>0\right\}  }$ is the probability that a service is completed in Server
$j$.

\textbf{Step two: Compute }$\Re_{2}=E\left[  r\left(  \mathbf{X}\left(
\eta_{2}\right)  \right)  \text{ }|\text{ }\mathbf{X}\left(  0\right)
=x,0<\eta_{1}<\eta_{2}<\cdots<\eta_{n}<t<\eta_{n+1}\right]  $

It is seen from (a) and (b) in Figure 4 that the Markov process $\left\{
\mathbf{X}\left(  t\right)  :t\geq0\right\}  $ transits to a state
$\mathbf{X}\left(  \eta_{2}\right)  $ from a state $\mathbf{X}\left(  \eta
_{1}\right)  $ in the set%
\[
\left\{  x-e_{j}:1\leq j\leq M\right\}  \cup\left\{  x+e_{k_{i}}:1\leq i\leq
M\right\}  ,
\]
hence we have%
\[
\mathbf{X}\left(  \eta_{2}\right)  =\left\{
\begin{array}
[c]{ll}%
x-e_{j}+e_{k_{m}}, & \text{if an arrival occurs in Server }k_{m}\text{ at time
}\eta_{2},\\
x+e_{k_{i}}+e_{k_{m}}, & \text{if an arrival occurs in Server }k_{m}\text{ at
time }\eta_{2},\\
x-e_{j}-e_{l} & \text{if a service is completed in Server }l\text{ at time
}\eta_{2},\\
x+e_{k_{i}}-e_{l} & \text{if a service is completed in Server }l\text{ at time
}\eta_{2},
\end{array}
\right.
\]
thus we have%
\begin{align*}
\mathbf{X}\left(  \eta_{2}\right)   &  \in\left\{  x-e_{j}+e_{k_{m}}:1\leq
j,m\leq M\right\}  \cup\left\{  x+e_{k_{i}}+e_{k_{m}}:1\leq i,m\leq M\right\}
\\
&  \cup\left\{  x-e_{j}-e_{l}:1\leq j,l\leq M\right\}  \cup\left\{
x+e_{k_{i}}-e_{l}:1\leq i,l\leq M\right\}  .
\end{align*}
Based on the above analysis, it is seen from (a) and (b) in Figure 4 that%
\begin{align}
\Re_{2}=  &  \left\{  \sum_{m=1}^{M}\sum_{j=1}^{M}r\left(  x-e_{j}+e_{k_{m}%
}\right)  \cdot b^{\left(  j\right)  }\mathbf{1}_{\left\{  x_{j}>0\right\}
}\cdot ak\left(  M,k_{m},d\right)  \right. \nonumber\\
&  +\left.  \sum_{m=1}^{M}\sum_{i=1}^{M}r\left(  x+e_{k_{i}}+e_{k_{m}}\right)
\cdot ak\left(  M,k_{i},d\right)  \cdot ak\left(  M,k_{m},d\right)  \right\}
\nonumber\\
&  +\left\{  \sum_{l=1}^{M}\sum_{j=1}^{M}r\left(  x-e_{j}-e_{l}\right)  \cdot
b^{\left(  j\right)  }\mathbf{1}_{\left\{  x_{j}>0\right\}  }\cdot b^{\left(
l\right)  }\mathbf{1}_{\left\{  \left(  x-e_{j}\right)  _{l}>0\right\}
}\right. \nonumber\\
&  +\left.  \sum_{l=1}^{M}\sum_{i=1}^{M}r\left(  x+e_{k_{i}}-e_{l}\right)
\cdot ak\left(  M,k_{i},d\right)  \cdot b^{\left(  l\right)  }\mathbf{1}%
_{\left\{  \left(  x+e_{k_{i}}\right)  _{l}>0\right\}  }\right\}  .
\label{Eq4.3}%
\end{align}

\textbf{Step three: Compute }$\Re_{k}=E\left[  r\left(  \mathbf{X}\left(
\eta_{k}\right)  \right)  \text{ }|\text{ }\mathbf{X}\left(  0\right)
=x,0<\eta_{1}<\eta_{2}<\cdots<\eta_{n}<t<\eta_{n+1}\right]  $ for $3\leq k\leq n$

From the above two special computations, here we will further develop the
event-driven technique to calculate the conditional mean of the random reward function.

For the general term $\Re_{k}$, our computation is more complicated than that
in the above two special cases. To that end, we need to introduce some
notation to record the order number of the server with either an arrival or a
service completion at each of the state jump points $\eta_{k}$ for
$k=1,2,\ldots,n$. Observing the two expressions (\ref{Eq4.2}) and
(\ref{Eq4.3}), the order numbers of the servers need to relate to the state
jump points $\eta_{k}$ for $k=1,2,\ldots,n$. For simplicity of description, it
is necessary to list some notation in Table 1, the purpose of which is to
express the state jump points and associated useful information.

\begin{table}[ptb]
\caption{The order number of servers with either an arrival or a service
completion}%
\renewcommand{\arraystretch}{1.3}  \centering
\begin{tabular}
[c]{|c|c|c|}\hline
\bfseries  State jump points & \bfseries  Server number by arrival &
\bfseries  Server number by service \bfseries \\\hline
$\eta_{1}$ & $k_{i_{1}}$ & $j_{1}$\\\hline
$\eta_{2}$ & $k_{i_{2}}$ & $j_{2}$\\\hline
$\vdots$ & $\vdots$ & $\vdots$\\\hline
$\eta_{n}$ & $k_{i_{n}}$ & $j_{n}$\\\hline
\end{tabular}
\end{table}

For simplification of description, when deriving some conditional means
involved below, we introduce a convention notation: $E_{Y}\left[  X\right]
=E\left[  E\left[  X|Y\right]  \right]  $ (that is, a deterministic value),
where $X$ and $Y$ are two random variables.

From Steps one and two, it is easy to see that $\Re_{k}$ depends on the $k$
successive samples for the states $\mathbf{X}\left(  \eta_{m}\right)  $ for
$m=1,2,\ldots,k-1$. To describe the states $\mathbf{X}\left(  \eta_{k}%
^{-}\right)  $, we express the successive state jumps as follows:
$\mathbf{X}\left(  0\right)  \underset{\rightarrow}{\times}\mathbf{X}\left(
\eta_{1}\right)  \underset{\rightarrow}{\times}\mathbf{X}\left(  \eta
_{2}\right)  \underset{\rightarrow}{\times}\cdots\underset{\rightarrow}%
{\times}\mathbf{X}\left(  \eta_{k-1}\right)  $, where $A\underset{\rightarrow
}{\times}B$ denote the Cartesian product from the set $A$ to the set $B$.
Since $\mathbf{X}\left(  \eta_{k}^{-}\right)  =\mathbf{X}\left(  \eta
_{k-1}\right)  $ and our computation depends on the $k-1$ successive samples
for the states $\mathbf{X}\left(  \eta_{m}\right)  $ for $m=1,2,\ldots,k-1$,
we set $\mathbf{X}\left(  \eta_{k}^{-}\right)  =\mathbf{X}\left(  0\right)
\underset{\rightarrow}{\times}\mathbf{X}\left(  \eta_{1}\right)
\underset{\rightarrow}{\times}\mathbf{X}\left(  \eta_{2}\right)
\underset{\rightarrow}{\times}\cdots\underset{\rightarrow}{\times}%
\mathbf{X}\left(  \eta_{k-1}\right)  $, hence the first $k-1$ samples
$\mathbf{X}\left(  0\right)  \underset{\rightarrow}{\times}\mathbf{X}\left(
\eta_{1}\right)  \underset{\rightarrow}{\times}\mathbf{X}\left(  \eta
_{2}\right)  \underset{\rightarrow}{\times}\cdots\underset{\rightarrow}%
{\times}\mathbf{X}\left(  \eta_{k-1}\right)  $ is used to record our previous
computational process. Therefore, we obtain%
\[
\Re_{k}=E_{\mathbf{X}\left(  0\right)  \underset{\rightarrow}{\times
}\mathbf{X}\left(  \eta_{1}\right)  \underset{\rightarrow}{\times}%
\mathbf{X}\left(  \eta_{2}\right)  \underset{\rightarrow}{\times}%
\cdots\underset{\rightarrow}{\times}\mathbf{X}\left(  \eta_{k-1}\right)
}\left[  r\left(  \mathbf{X}\left(  \eta_{k}\right)  \right)  \right]  .
\]
In this case, we need to represent the initial state $\mathbf{X}\left(
0\right)  \underset{\rightarrow}{\times}\mathbf{X}\left(  \eta_{1}\right)
\underset{\rightarrow}{\times}\mathbf{X}\left(  \eta_{2}\right)
\underset{\rightarrow}{\times}\cdots\underset{\rightarrow}{\times}%
\mathbf{X}\left(  \eta_{k-1}\right)  $ by means of $\mathbf{X}\left(  \eta
_{m}\right)  \in\left\{  \bullet-e_{j_{m}}:1\leq j_{m}\leq M\right\}
\cup\left\{  \bullet+e_{k_{i_{m}}}:1\leq i_{m}\leq M\right\}  $ for $1\leq
m\leq k-1$, thus we have%
\[
\mathbf{X}\left(  0\right)  \underset{\rightarrow}{\times}\mathbf{X}\left(
\eta_{1}\right)  \underset{\rightarrow}{\times}\mathbf{X}\left(  \eta
_{2}\right)  \underset{\rightarrow}{\times}\cdots\underset{\rightarrow}%
{\times}\mathbf{X}\left(  \eta_{k-1}\right)  \in\Theta^{\left(  k-1\right)
},
\]
where%
\[
\Theta^{\left(  k-1\right)  }=\Theta_{0}\times\Theta_{1}\times\Theta_{2}%
\times\cdots\times\Theta_{k-1},
\]%
\begin{align*}
\Theta_{0}  &  =\left\{  x\right\} \\
\Theta_{1}  &  =\left\{  \bullet-e_{j_{1}}:1\leq j_{1}\leq M\right\}
\cup\left\{  \bullet+e_{k_{i_{1}}}:1\leq i_{1}\leq M\right\}  ,\\
\Theta_{2}  &  =\left\{  \bullet-e_{j_{2}}:1\leq j_{2}\leq M\right\}
\cup\left\{  \bullet+e_{k_{i_{2}}}:1\leq i_{2}\leq M\right\}  ,\\
&  \vdots\text{
\ \ \ \ \ \ \ \ \ \ \ \ \ \ \ \ \ \ \ \ \ \ \ \ \ \ \ \ \ \ \ \ \ \ \ }%
\vdots\\
\Theta_{k-1}  &  =\left\{  \bullet-e_{j_{k-1}}:1\leq j_{k-1}\leq M\right\}
\cup\left\{  \bullet+e_{k_{i_{k-1}}}:1\leq i_{k-1}\leq M\right\}  .
\end{align*}

To understand the elements in the set $\Theta^{\left(  k-1\right)  }$, we need
the Cartesian product as follows:%
\[
\left\{  A,B\right\}  \times\left\{  C,D\right\}  =\left\{  A\times C,A\times
D,B\times C,B\times D\right\}  ,
\]
where $A,B,C,D$ are four sets with finite elements.

In the set $\Theta^{\left(  k-1\right)  }$, the $k$ elements are successively
taken from the subsets $\Theta_{0},\Theta_{1},\Theta_{2},\ldots,$
$\Theta_{k-1}$, for example, $x\in\Theta_{0}$, $\bullet-e_{j_{1}}\in\Theta
_{1}$, $\bullet+e_{k_{i_{2}}}\in\Theta_{2}$, \ldots, $\bullet-e_{j_{k-2}}%
\in\Theta_{k-2}$, $\bullet+e_{k_{i_{k-1}}}\in\Theta_{k-1}$. For the successive
$k$ elements, we have a simple computation through the following convention%
\[
\left\{  x\right\}  \left\{  \bullet-e_{j_{1}}\right\}  =x-e_{j_{1}},
\]%
\[
\left\{  x\right\}  \left\{  \bullet-e_{j_{1}}\right\}  \left\{
\bullet+e_{k_{i_{2}}}\right\}  =x-e_{j_{1}}+e_{k_{i_{2}}},
\]%
\[
\cdots\cdots\cdots
\]%
\[
\left\{  x\right\}  \left\{  \bullet-e_{j_{1}}\right\}  \left\{
\bullet+e_{k_{i_{2}}}\right\}  \cdots\left\{  \bullet-e_{j_{k-2}}\right\}
\left\{  \bullet+e_{k_{i_{k-1}}}\right\}  =x-e_{j_{1}}+e_{k_{i_{2}}}%
\cdots-e_{j_{k-2}}+e_{k_{i_{k-1}}}.
\]
Based on this, we can easily give a sample of the initial state $\mathbf{X}%
\left(  0\right)  \underset{\rightarrow}{\times}\mathbf{X}\left(  \eta
_{1}\right)  \underset{\rightarrow}{\times}\mathbf{X}\left(  \eta_{2}\right)
\underset{\rightarrow}{\times}\cdots\underset{\rightarrow}{\times}%
\mathbf{X}\left(  \eta_{k-1}\right)  $ in the set $\Theta^{\left(  k-1\right)
}$.

Now, we compute the conditional mean $E_{\mathbf{X}\left(  0\right)
\underset{\rightarrow}{\times}\mathbf{X}\left(  \eta_{1}\right)
\underset{\rightarrow}{\times}\mathbf{X}\left(  \eta_{2}\right)  \text{
\ }\underset{\rightarrow}{\times}\cdots\underset{\rightarrow}{\times
}\mathbf{X}\left(  \eta_{k-1}\right)  }\left[  r\left(  \mathbf{X}\left(
\eta_{k}\right)  \right)  \right]  $ by means of an iterative algorithm as follows:

(a) For $m=1$, we have%
\[
E_{\mathbf{X}\left(  0\right)  }\left[  r\left(  \mathbf{X}\left(  \eta
_{1}\right)  \right)  \right]  =\sum_{i=1}^{M}ak\left(  M,k_{i},d\right)
r\left(  x+e_{k_{i}}\right)  +\sum_{i=1}^{M}b^{\left(  j\right)  }%
\mathbf{1}_{\left\{  x_{j}>0\right\}  }r\left(  x-e_{j}\right)  .
\]

(b) For $m=2$, we have%
\begin{align*}
E_{\mathbf{X}\left(  0\right)  \underset{\rightarrow}{\times}\mathbf{X}\left(
\eta_{1}\right)  }\left[  r\left(  \mathbf{X}\left(  \eta_{2}\right)  \right)
\right]  =  &  \left\{  \sum_{m=1}^{M}\sum_{j=1}^{M}r\left(  x-e_{j}+e_{k_{m}%
}\right)  \cdot b^{\left(  j\right)  }\mathbf{1}_{\left\{  x_{j}>0\right\}
}\cdot ak\left(  M,k_{m},d\right)  \right. \\
&  +\left.  \sum_{m=1}^{M}\sum_{i=1}^{M}r\left(  x+e_{k_{i}}+e_{k_{m}}\right)
\cdot ak\left(  M,k_{i},d\right)  \cdot ak\left(  M,k_{m},d\right)  \right\}
\\
&  +\left\{  \sum_{l=1}^{M}\sum_{j=1}^{M}r\left(  x-e_{j}-e_{l}\right)  \cdot
b^{\left(  j\right)  }\mathbf{1}_{\left\{  x_{j}>0\right\}  }\cdot b^{\left(
l\right)  }\mathbf{1}_{\left\{  \left(  x-e_{j}\right)  _{l}>0\right\}
}\right. \\
&  +\left.  \sum_{l=1}^{M}\sum_{i=1}^{M}r\left(  x+e_{k_{i}}-e_{l}\right)
\cdot ak\left(  M,k_{i},d\right)  \cdot b^{\left(  l\right)  }\mathbf{1}%
_{\left\{  \left(  x+e_{k_{i}}\right)  _{l}>0\right\}  }\right\}  ,
\end{align*}

(c) For $m=k\geq3$, we take an element $y_{k-1}\in\Theta^{\left(  k-1\right)
}$, then%
\begin{align}
E_{y_{k-1}}\left[  r\left(  \mathbf{X}\left(  \eta_{k}\right)  \right)
\right]  =  &  \left[  \sum_{i_{k}=1}^{M}r\left(  y_{k-1}+e_{k_{i_{k}}%
}\right)  f\left(  y_{k-1}\right)  \cdot ak\left(  M,k_{i_{k}},d\right)
\right. \nonumber\\
&  \left.  +\sum_{l=1}^{M}r\left(  y_{k-1}-e_{l}\right)  f\left(
y_{k-1}\right)  \cdot b^{\left(  l\right)  }\mathbf{1}_{\left\{  \left(
y_{k-1}\right)  _{l}>0\right\}  }\right]  , \label{Eq4-5}%
\end{align}
and $f\left(  y_{k-1}\right)  $ is the probability that the state $y_{k-1}$
occurs. It is necessary to provide some interpretation for the probability
$f\left(  y_{k-1}\right)  $ by means of the following three examples:

\textbf{(c-1)} $f\left(  y_{0}\right)  =1$ due to $y_{0}=x$.

\textbf{(c-2)} If $y_{1}=$ $x-e_{j_{1}}$, then $f\left(  y_{1}\right)
=b^{\left(  j_{1}\right)  }\mathbf{1}_{\left\{  x_{j_{1}}>0\right\}  }$; If
$y_{1}=$ $x+e_{k_{i_{1}}}$, then $f\left(  y_{1}\right)  =ak\left(
M,k_{i_{1}},d\right)  $.

\textbf{(c-3)} If $y_{2}=$ $x-e_{j_{1}}-e_{j_{2}}$, then $f\left(
y_{2}\right)  =b^{\left(  j_{1}\right)  }\mathbf{1}_{\left\{  x_{j_{1}%
}>0\right\}  }b^{\left(  j_{2}\right)  }\mathbf{1}_{\left\{  \left(
x-e_{j_{1}}\right)  _{j_{2}}>0\right\}  }$; if $y_{2}=$ $x+e_{k_{i_{1}}%
}-e_{j_{2}}$, then $f\left(  y_{2}\right)  =ak\left(  M,k_{i_{1}},d\right)
b^{\left(  j_{2}\right)  }\mathbf{1}_{\left\{  \left(  x+e_{k_{i_{1}}}\right)
_{j_{2}}>0\right\}  }$; and the other two can similarly be computed and both
of them are omitted here.

Note that $y_{k-1}\in\Theta^{\left(  k-1\right)  }$, using (\ref{Eq4-5}) we
obtain%
\begin{align}
&  E_{\mathbf{X}\left(  0\right)  \underset{\rightarrow}{\times}%
\mathbf{X}\left(  \eta_{1}\right)  \underset{\rightarrow}{\times}%
\mathbf{X}\left(  \eta_{2}\right)  \underset{\rightarrow}{\times}%
\cdots\underset{\rightarrow}{\times}\mathbf{X}\left(  \eta_{k-1}\right)
}\left[  r\left(  \mathbf{X}\left(  \eta_{k}\right)  \right)  \right]
\nonumber\\
=  &  \sum_{y_{k-1}\in\Theta^{\left(  k-1\right)  }}\left[  \sum_{i_{k}=1}%
^{M}r\left(  y_{k-1}+e_{k_{i_{k}}}\right)  f\left(  y_{k-1}\right)  \cdot
ak\left(  M,k_{i_{k}},d\right)  \right. \nonumber\\
&  \left.  +\sum_{l=1}^{M}r\left(  y_{k-1}-e_{l}\right)  f\left(
y_{k-1}\right)  \cdot b^{\left(  l\right)  }\mathbf{1}_{\left\{  \left(
y_{k}\right)  _{l}>0\right\}  }\right]  . \label{Eq4.6}%
\end{align}

Now, we further discuss the key computation of $f\left(  y_{k-1}\right)  $
whose purpose is to provide some new highlight on the calculation program.

Intuitively, the set of jump states: $\mathbf{X}\left(  \eta_{1}\right)
\underset{\rightarrow}{\times}\mathbf{X}\left(  \eta_{2}\right)
\underset{\rightarrow}{\times}\cdots\underset{\rightarrow}{\times}%
\mathbf{X}\left(  \eta_{k}\right)  $, can be decomposed into two subsets: One
for an arrival and another for a service completion. Based on this, we record
the order numbers for either the arrivals or the service completions, for
example, if $\mathbf{X}\left(  \eta_{m}\right)  $ occurs at an arrival, then
we record the order number as $V_{m}$; while if $\mathbf{X}\left(  \eta
_{m}\right)  $ occurs at a service completion, then we record the order number
as $W_{m}$. Therefore, the set of the order numbers is given by%
\[
\left\{  1,2,3,\ldots,k\right\}  =\left\{  V_{i_{1}},V_{i_{2}},V_{i_{3}%
},\ldots,V_{i_{p}}\right\}  \cup\left\{  W_{j_{1}},W_{j_{2}},W_{j_{3}}%
,\ldots,W_{j_{k-p}}\right\}  ,
\]
where $0\leq p\leq k$. Specifically, if $p=0$, then the set of the order
numbers only contains the service completions; while if $p=k$, then the set of
the order numbers only contains the arrivals.

Based on the two subsets $\left\{  V_{i_{1}},V_{i_{2}},V_{i_{3}}%
,\ldots,V_{i_{p}}\right\}  $ and $\left\{  W_{j_{1}},W_{j_{2}},W_{j_{3}%
},\ldots,W_{j_{k-p}}\right\}  $, we obtain%
\[
a^{p}\underset{m=1}{\overset{p}{\Pi}}k(M,k_{i_{V_{i_{m}}}},d)\cdot
\overset{k-p}{\underset{h=1}{\Pi}}b^{(j_{W_{h}})}\mathbf{1}_{\left\{  \left(
x-\overset{h-1}{\underset{m=1}{\sum}}e_{j_{W_{m}}}+\underset{V_{s}\leq
W_{h}-1}{\sum}e_{k_{i_{V_{s}}}}\right)  _{j_{W_{h}}}>0\right\}  }\cdot
r\left(  x+\underset{m=1}{\overset{p}{\sum}}e_{k_{i_{V_{i_{m}}}}}%
-\underset{h=1}{\overset{k-p}{\sum}}e_{j_{W_{h}}}\right)  ,
\]
where we have some convention on $\underset{m=1}{\overset{0}{\Pi}}\bullet=1$
and $\underset{m=1}{\overset{0}{\sum}}\bullet=0$, and notice that
$k_{i_{V_{i_{m}}}}$ depends on the state $y_{\left(  V_{i_{m}}-1\right)  }$.
Thus we obtain%
\begin{align}
&  E_{\mathbf{X}\left(  0\right)  \underset{\rightarrow}{\times}%
\mathbf{X}\left(  \eta_{1}\right)  \underset{\rightarrow}{\times}%
\mathbf{X}\left(  \eta_{2}\right)  \underset{\rightarrow}{\times}%
\cdots\underset{\rightarrow}{\times}\mathbf{X}\left(  \eta_{k-1}\right)
}\left[  r\left(  \mathbf{X}\left(  \eta_{k}\right)  \right)  \right]
\nonumber\\
&  =\underset{p=0}{\overset{k}{\sum}}\sum_{i_{V_{i_{1}}}=1}^{M}\cdots
\sum_{i_{V_{i_{p}}}=1}^{M}\underset{j_{W_{1}}=1}{\overset{M}{\sum}}%
\cdots\underset{j_{W_{k-p}}=1}{\overset{M}{\sum}}a^{p}\underset{m=1}%
{\overset{p}{\Pi}}k(M,k_{i_{V_{i_{m}}}},d)\nonumber\\
&  \times\text{ }\overset{k-p}{\underset{h=1}{\Pi}}b^{(j_{W_{h}})}%
\mathbf{1}_{\left\{  \left(  x-\overset{h-1}{\underset{m=1}{\sum}}e_{j_{W_{m}%
}}+\underset{V_{s}\leq W_{h}-1}{\sum}e_{k_{i_{V_{s}}}}\right)  _{j_{W_{h}}%
}>0\right\}  }\cdot r\left(  x+\underset{m=1}{\overset{p}{\sum}}%
e_{k_{i_{V_{i_{m}}}}}-\underset{h=1}{\overset{k-p}{\sum}}e_{j_{W_{h}}}\right)
. \label{Eq4.7}%
\end{align}

Similarly, from the two subsets $\left\{  V_{i_{1}},V_{i_{2}},V_{i_{3}}%
,\ldots,V_{i_{p}}\right\}  $ and $\left\{  W_{j_{1}},W_{j_{2}},W_{j_{3}%
},\ldots,W_{j_{k-p}}\right\}  $ we obtain%
\[
f\left(  y_{k}\right)  =a^{p}\underset{m=1}{\overset{p}{\Pi}}%
k(M,k_{i_{V_{i_{m}}}},d)\cdot\overset{k-p}{\underset{h=1}{\Pi}}b^{(j_{W_{h}}%
)}\mathbf{1}_{\left\{  \left(  x-\overset{h-1}{\underset{m=1}{\sum}%
}e_{j_{W_{m}}}+\underset{V_{s}\leq W_{h}-1}{\sum}e_{k_{i_{V_{s}}}}\right)
_{j_{W_{h}}}>0\right\}  }.
\]

In the remainder of this section, we finally compute the conditional mean
$E\left[  \Phi\left(  t\right)  \text{ }|\text{ }\mathbf{X}\left(  0\right)
=x\right]  $ of the stochastic integral $\Phi\left(  t\right)  $ according to
the above steps one to three.

It follows from (\ref{Eq4.1}) that%
\begin{align}
E\left[  \Phi\left(  t\right)  \text{ }|\text{ }\mathbf{X}\left(  0\right)
=x\right]   &  =r\left(  x\right)  te^{-\omega t}+\sum_{n=1}^{\infty
}e^{-\omega t}\frac{\left(  \omega t\right)  ^{n}}{n!}\cdot\frac{t}%
{n+1}\nonumber\\
&  \times\left\{  r\left(  x\right)  +\sum_{k=1}^{n}E_{\mathbf{X}\left(
0\right)  \underset{\rightarrow}{\times}\mathbf{X}\left(  \eta_{1}\right)
\underset{\rightarrow}{\times}\mathbf{X}\left(  \eta_{2}\right)
\underset{\rightarrow}{\times}\cdots\underset{\rightarrow}{\times}%
\mathbf{X}\left(  \eta_{k-1}\right)  }\left[  r\left(  \mathbf{X}\left(
\eta_{k}\right)  \right)  \right]  \right\}  , \label{Eq4.8}%
\end{align}
where $E_{\mathbf{X}\left(  0\right)  \underset{\rightarrow}{\times}%
\mathbf{X}\left(  \eta_{1}\right)  \underset{\rightarrow}{\times}%
\mathbf{X}\left(  \eta_{2}\right)  \underset{\rightarrow}{\times}%
\cdots\underset{\rightarrow}{\times}\mathbf{X}\left(  \eta_{k-1}\right)
}\left[  r\left(  \mathbf{X}\left(  \eta_{k}\right)  \right)  \right]  $ is
given in (\ref{Eq4.6}) or (\ref{Eq4.7}).

\section{A Markov Discounted Reward Process}

In this section, we provide an effective method for computing the mean of the
discounted random reward function in an infinite time interval. Based on this,
we give a simple discussion on optimal criterions for designing the
supermarket model with different servers.

In the infinite time interval $[0,+\infty)$, it is possible that $E\left[
\Phi\left(  +\infty\right)  \text{ }|\text{ }\mathbf{X}\left(  0\right)
=x\right]  =+\infty$. To avoid the infinite case, the random reward function
is always taken as a discounted reward. Notice that $r\left(  x\right)  $ is a
reward value of the $M$-dimensional Markov process $\left\{  \mathbf{X}\left(
t\right)  :t\geq0\right\}  $ at state $x\in\Omega$, we define a discounted
random reward function as%
\begin{equation}
\Psi\left(  \beta\right)  =\int_{0}^{+\infty}e^{-\beta t}r\left(
\mathbf{X}\left(  t\right)  \right)  \text{d}t, \label{Equ5-1}%
\end{equation}
where $\beta\geq0$ is a discounted rate, and the discounted factor $e^{-\beta
t}$ guarantees that $\Psi\left(  \beta\right)  $ is finite a.s..

If $\Psi\left(  0\right)  $ is finite a.s., then $\Psi\left(  0\right)
=E\left[  \Phi\left(  +\infty\right)  \right]  $ is an ordinary
(non-discounted) random reward function, as studied in Section 4 with
$t\rightarrow+\infty$.

Now, we provide a segmented stochastic integral for expressing the random
reward function $\Psi\left(  \beta\right)  $, this will be useful in our
following computation.

Let $\eta_{1},\eta_{2},\eta_{3},\ldots$ be the successive state jump points of
the $M$-dimensional Markov process $\left\{  \mathbf{X}\left(  t\right)
:t\geq0\right\}  $ in the time interval $[0,+\infty)$, it is clear from the
Poisson or exponential assumptions that%
\[
0<\eta_{1}<\eta_{2}<\eta_{3}<\cdots.
\]
At the same time, the sequence: $\eta_{1}$, $\eta_{n+1}-\eta_{n}$ for $n\geq
1$, is i.d.d. and exponential with mean $1/\omega$. Note that the case with
the time interval $[0,+\infty)$ is different from that in Section 4 with
respect to analysis of the uniform distributions.

Note that%
\[
\lbrack0,+\infty)=[0,\eta_{1}^{-})\cup\lbrack\eta_{1},\eta_{2}^{-})\cup
\lbrack\eta_{2},\eta_{3}^{-})\cup\cdots,
\]
it follows from (\ref{Equ5-1})\ that%
\begin{equation}
\Psi\left(  \beta\right)  =\int_{0}^{\eta_{1}^{-}}e^{-\beta t}r\left(
\mathbf{X}\left(  t\right)  \right)  \text{d}t+\sum_{j=1}^{\infty}\int
_{\eta_{j}}^{\eta_{j+1}^{-}}e^{-\beta t}r\left(  \mathbf{X}\left(  t\right)
\right)  \text{d}t. \label{Equ5-2}%
\end{equation}
Thus we obtain
\begin{align}
E\left[  \Psi\left(  \beta\right)  \text{ }|\text{ }\mathbf{X}\left(
0\right)  =x\right]   &  =E_{x}\left[  \int_{0}^{\eta_{1}^{-}}e^{-\beta
t}r\left(  \mathbf{X}\left(  t\right)  \right)  \text{d}t\right]  +\sum
_{k=1}^{\infty}E_{\mathbf{X}\left(  \eta_{k}^{-}\right)  }\left[  \int
_{\eta_{k}}^{\eta_{k+1}^{-}}e^{-\beta t}r\left(  \mathbf{X}\left(  t\right)
\right)  \text{d}t\right] \nonumber\\
&  =r\left(  x\right)  E_{x}\left[  \int_{0}^{\eta_{1}^{-}}e^{-\beta
t}\text{d}t\right]  +\sum_{k=1}^{\infty}E_{\mathbf{X}\left(  \eta_{k}%
^{-}\right)  }\left[  r\left(  \eta_{k}\right)  \int_{\eta_{k}}^{\eta
_{k+1}^{-}}e^{-\beta t}\text{d}t\right]  . \label{Equ5-3}%
\end{align}
Note that our following computation shows that $E\left[  \Psi\left(
\beta\right)  \text{ }|\text{ }\mathbf{X}\left(  0\right)  =x\right]  $ is not
about the taken sequence $\left\{  \eta_{k}:k\geq1\right\}  $.

Since $r\left(  \mathbf{X}\left(  t\right)  \right)  =r\left(  x\right)  $ for
$t\in\lbrack0,\eta_{1}^{-})$ and $r\left(  \mathbf{X}\left(  t\right)
\right)  =r\left(  \eta_{k}\right)  $ for $t\in\lbrack\eta_{k},\eta_{k+1}%
^{-})$, we need to compute $E\left[  \int_{0}^{\eta_{1}^{-}}e^{-\beta
t}\text{d}t\right]  $ and $E\left[  \int_{\eta_{k}}^{\eta_{k+1}^{-}}e^{-\beta
t}\text{d}t\right]  $ for $k\geq1$.

It is easy to check that%
\begin{equation}
E\left[  \int_{0}^{\eta_{1}^{-}}e^{-\beta t}\text{d}t\right]  =\frac{1}%
{\beta+\lambda+\mu_{1}+\mu_{2}+\cdots+\mu_{M}}. \label{Equ5-4}%
\end{equation}

To compute $E\left[  \int_{\eta_{k}}^{\eta_{k+1}^{-}}e^{-\beta t}%
\text{d}t\right]  $, let the random variable $\Gamma$ be exponential with
parameter $\lambda+\mu_{1}+\mu_{2}+\cdots+\mu_{M}$. Then we have%
\begin{equation}
E\left[  \int_{\eta_{k}}^{\eta_{k+1}^{-}}e^{-\beta t}\text{d}t\right]
=E^{\left(  \Gamma\right)  }\left[  E^{\left(  \eta_{k}\right)  }\left[
\int_{\eta_{k}}^{\eta_{k}+\Gamma}e^{-\beta t}\text{d}t\right]  \right]  ,
\label{Equ5-4-1}%
\end{equation}
where $E^{\left(  Y\right)  }\left[  \bullet\right]  $ denote such a mean with
respect to the random variable $Y$. It is clear that $\eta_{k}$ is a random
variable with the Erlang distribution of order $k$ as follows:%
\[
P\left\{  \eta_{k}\leq y\right\}  =1-\exp\left\{  -\left(  \lambda+\mu_{1}%
+\mu_{2}+\cdots+\mu_{M}\right)  y\right\}  \sum_{j=0}^{k-1}\frac{\left[
\left(  \lambda+\mu_{1}+\mu_{2}+\cdots+\mu_{M}\right)  y\right]  ^{j}}{j!}.
\]
Hence it follows from (\ref{Equ5-4-1}) that%
\begin{equation}
E\left[  \int_{\eta_{k}}^{\eta_{k+1}^{-}}e^{-\beta t}\text{d}t\right]
=\int_{0}^{+\infty}\int_{0}^{+\infty}\int_{y}^{y+x}e^{-\beta t}\text{d}%
t\text{d}P\left\{  \Gamma\leq x\right\}  \text{d}P\left\{  \eta_{k}\leq
y\right\}  . \label{Equ5-4-2}%
\end{equation}

Based on (\ref{Equ5-4-1}) and (\ref{Equ5-4-2}), we set%
\[
\theta_{0}\left(  \beta\right)  =E\left[  \int_{0}^{\eta_{1}^{-}}e^{-\beta
t}\text{d}t\right]
\]
and for $k\geq1$%
\[
\theta_{k}\left(  \beta\right)  =E\left[  \int_{\eta_{k}}^{\eta_{k+1}^{-}%
}e^{-\beta t}\text{d}t\right]  .
\]
Note that the sequence $\left\{  \theta_{n}\left(  \beta\right)
:n\geq0\right\}  $ can explicitly be determined by (\ref{Equ5-4-1}) and
(\ref{Equ5-4-2}), although we omit some computational details.

It is easy to check that%
\[
E_{\mathbf{X}\left(  0\right)  }\left[  \int_{0}^{\eta_{1}^{-}}e^{-\beta
t}r\left(  \mathbf{X}\left(  t\right)  \right)  \text{d}t\right]
=\frac{r\left(  x\right)  }{\beta+\lambda+\mu_{1}+\mu_{2}+\cdots+\mu_{M}%
}=\theta_{0}\left(  \beta\right)  r\left(  x\right)  .
\]

Now, we compute $E_{\mathbf{X}\left(  \eta_{k}^{-}\right)  }\left[  \int
_{\eta_{k}}^{\eta_{k+1}^{-}}e^{-\beta t}r\left(  \mathbf{X}\left(  t\right)
\right)  \text{d}t\right]  $ by a similar method given in (\ref{Eq4.6}) as
follows:%
\begin{align}
&  E_{\mathbf{X}\left(  0\right)  \underset{\rightarrow}{\times}%
\mathbf{X}\left(  \eta_{1}\right)  \underset{\rightarrow}{\times}%
\mathbf{X}\left(  \eta_{2}\right)  \underset{\rightarrow}{\times}%
\cdots\underset{\rightarrow}{\times}\mathbf{X}\left(  \eta_{k-1}\right)
}\left[  \int_{\eta_{k}}^{\eta_{k+1}^{-}}e^{-\beta t}r\left(  \mathbf{X}%
\left(  t\right)  \right)  \text{d}t\right] \nonumber\\
=  &  \theta_{k}\left(  \beta\right)  \sum_{y_{k-1}\in\Theta^{\left(
k-1\right)  }}\left[  \sum_{i_{k}=1}^{M}r\left(  y_{k-1}+e_{k_{i_{k}}}\right)
f\left(  y_{k-1}\right)  \cdot ak\left(  M,k_{i_{k}},d\right)  \right.
\nonumber\\
&  \left.  +\sum_{l=1}^{M}r\left(  y_{k-1}-e_{l}\right)  f\left(
y_{k-1}\right)  \cdot b^{\left(  l\right)  }\mathbf{1}_{\left\{  \left(
y_{k}\right)  _{l}>0\right\}  }\right]  . \label{Equ5-5}%
\end{align}
It follows from (\ref{Equ5-3}), (\ref{Equ5-4}) and (\ref{Equ5-5}) that%
\begin{align}
E\left[  \Psi\left(  \beta\right)  \text{ }|\text{ }\mathbf{X}\left(
0\right)  =x\right]   &  =\theta_{0}\left(  \beta\right)  r\left(  x\right)
+\sum_{k=1}^{\infty}\theta_{k}\left(  \beta\right)  \sum_{y_{k-1}\in
\Theta^{\left(  k-1\right)  }}\left[  \sum_{i_{k}=1}^{M}r\left(
y_{k-1}+e_{k_{i_{k}}}\right)  f\left(  y_{k-1}\right)  \right. \nonumber\\
&  \left.  \times ak\left(  M,k_{i_{k}},d\right)  +\sum_{l=1}^{M}r\left(
y_{k-1}-e_{l}\right)  f\left(  y_{k-1}\right)  \cdot b^{\left(  l\right)
}\mathbf{1}_{\left\{  \left(  y_{k}\right)  _{l}>0\right\}  }\right]  .
\label{Equ5-6}%
\end{align}
It is seen from (\ref{Equ5-6}) that $E\left[  \Psi\left(  \beta\right)  \text{
}|\text{ }\mathbf{X}\left(  0\right)  =x\right]  $ is discounted by the
$\beta$-sequence $\left\{  \theta_{n}\left(  \beta\right)  :n\geq0\right\}  $,
which guarantees that $E\left[  \Psi\left(  \beta\right)  \text{ }|\text{
}\mathbf{X}\left(  0\right)  =x\right]  <+\infty$.

In the remainder of this section, we provide a simple discussion for optimal
design of the supermarket model with different servers. Specifically, such an
optimization may be realized through an event-driven technique with
performance simulation as well as perturbation realization, e.g., see Cao
\cite{Cao:2007} and Xia and Cao \cite{Xia:2012}.

To realize an optimal design, the parameters of this supermarket model can be
classified as three different groups: (1) The customer arrival parameters
$\lambda$; and $g_{1},g_{2},\ldots,g_{M}$. (2) The customer service parameters
$M$; $d$; $\mu_{1},\mu_{2},\ldots,\mu_{M}$. (3) The economic parameters
$r\left(  x\right)  $ for $x\in\Omega$. In general, the customer arrival
parameters are always fixed, given that the customer resource and environment
are fixed; while the economic parameters are chosen in order that performance
optimization of this supermarket model can be easy to be carried out. Based on
this, our optimal design is to focus on taking the optimal service parameters:
$M$; $d$; $\mu_{1},\mu_{2},\ldots,\mu_{M}$.

From a practical point of view of performance optimization, we take two
different reward values: $r_{\min}(x):=\Delta_{k_{1}}(x)$, and $r_{\max
}(x):=\Delta_{k_{M}}(x)$ for $x\in\Omega$, respectively. Thus, for $r\left(
x\right)  =r_{\min}(x)=\Delta_{k_{1}}(x)$ for $x\in\Omega$, we write%
\[
\mathbf{\Psi}\left(  \beta,r_{\min}\right)  =E\left[  \Psi\left(
\beta,r_{\min}\right)  \text{ }|\text{ }\mathbf{X}\left(  0\right)  =x\right]
;
\]
while for $r\left(  x\right)  =r_{\max}(x)=\Delta_{k_{M}}(x)$ for $x\in\Omega
$, we set%
\[
\mathbf{\Psi}\left(  \beta,r_{\max}\right)  =E\left[  \Psi\left(
\beta,r_{\max}\right)  \text{ }|\text{ }\mathbf{X}\left(  0\right)  =x\right]
.
\]

Based on $E\left[  \Psi\left(  \beta\right)  \text{ }|\text{ }\mathbf{X}%
\left(  0\right)  =x\right]  $, using an event-driven technique with
performance simulation as well as perturbation realization, we can obtain the
optimal decision parameters $M^{\ast}$; $d^{\ast}$; $\mu_{1}^{\ast},\mu
_{2}^{\ast},\ldots,\mu_{M}^{\ast}$ such that%
\begin{equation}
\mathbf{\Psi}^{\ast}\left(  \beta,r_{\min}\right)  =\max\left\{  \mathbf{\Psi
}\left(  \beta,r_{\min}\right)  \right\}  . \label{Opt-1}%
\end{equation}
Similarly, we can also give the optimal decision parameters $M^{\diamond}$;
$d^{\diamond}$; $\mu_{1}^{\diamond},\mu_{2}^{\diamond},\ldots,\mu
_{M}^{\diamond}$ such that%
\begin{equation}
\mathbf{\Psi}^{\diamond}\left(  \beta,r_{\max}\right)  =\min\left\{
\mathbf{\Psi}\left(  \beta,r_{\max}\right)  \right\}  . \label{Opt-2}%
\end{equation}
Furthermore, we can get the optimal decision parameters $M^{\bigtriangledown}%
$; $d^{\bigtriangledown}$; $\mu_{1}^{\bigtriangledown},\mu_{2}%
^{\bigtriangledown},\ldots,\mu_{M}^{\bigtriangledown}$ such that%
\begin{equation}
\mathbf{L}^{\bigtriangledown}\left(  \beta\right)  =\min\left\{  \mathbf{\Psi
}\left(  \beta,r_{\max}\right)  -\mathbf{\Psi}\left(  \beta,r_{\min}\right)
\right\}  . \label{Opt-3}%
\end{equation}

According to the above analysis, to design the supermarket model with
different servers, it is seen from Equations (\ref{Opt-1}) to (\ref{Opt-3})
that here we provide two optimal criterions as follows:

\textbf{Criterion one: }This supermarket model is better when choosing some
parameters such that $|\mathbf{\Psi}^{\diamond}\left(  \beta,r_{\max}\right)
-\mathbf{\Psi}^{\ast}\left(  \beta,r_{\min}\right)  |<\delta_{1}$ for a given
value $\delta_{1}>0$.

\textbf{Criterion two: }This supermarket model is better when choosing some
parameters such that $\mathbf{L}^{\bigtriangledown}\left(  \beta\right)
<\delta_{2}$ for a given value $\delta_{2}>0$.

In general, the two optimal criterions can easily be implemented by means of
the event-driven technique with performance simulation as well as perturbation
realization, e.g., see Cao \cite{Cao:2007} and Xia and Cao \cite{Xia:2012} for
more details.

\section{Performance Simulation}

In this section, we provide three simulation experiments whose purpose is to
simply discuss how the expected queue length of each server depends on some
key parameters: The choice number $d$, the service rate vector $\mu$ and the
probability vector $g$ of individual preference in the supermarket model with
different servers.

In the three simulation experiments, we take the server number $M=10$ and the
arrival rate $\lambda=10$.

\textbf{Experiment one:} In the supermarket model with different servers, we
take that the choice number $d=2$; the service rates of the $10$ servers are
listed as $\mu_{1}=1.1$, $\mu_{2}=1.2$, $\mu_{3}=1.3$, $\mu_{4}=1.4$, $\mu
_{5}=1.5$, $\mu_{6}=1.6$, $\mu_{7}=1.7$, $\mu_{8}=1.8$, $\mu_{9}=1.9$ and
$\mu_{10}=2.0$, respectively; the probabilities of individual preference for
the $10$ servers are given by $g_{1}=0.10$, $g_{2}=0.20$, $g_{3}=0.30$,
$g_{4}=0.05$, $g_{5}=0.05$, $g_{6}=0.02$, $g_{7}=0.10$, $g_{8}=0.03$,
$g_{9}=0.10$ and $g_{10}=0.05$, respectively. We simulate the expected
queueing length for each server by using the routine selection function%
\[
\Delta_{i}(x)=\frac{1+\frac{x_{i}}{\mu_{i}g_{i}}}{\underset{j=1}{\overset
{M}{\sum}}\left[  1+\frac{x_{j}}{\mu_{j}g_{j}}\right]  },\text{ \ }%
i=1,2,\ldots,10.
\]
The experimented results are shown in Table 1.

\begin{table}[ptb]
\caption{The expected queue lengths in the 10 servers}%
\renewcommand{\arraystretch}{1.3}  \centering
\begin{tabular}
[c]{|c|c|c|}\hline
\bfseries {\bf Server number}  & \bfseries {\bf Expected queue lengths}  &
\\\hline
One & 0.6834 & \\\hline
Two & 0.9454 & \\\hline
Three & 1.0440 & \\\hline
Four & 0.4318 & \\\hline
Five & 0.4234 & \\\hline
Six & 0.2894 & \\\hline
Seven & 0.4864 & \\\hline
Eight & 0.2793 & \\\hline
Nine & 0.4319 & \\\hline
Ten & 0.2640 & \\\hline
\end{tabular}
\end{table}

\textbf{Experiment two:} This experiment takes the different parameters: only
the 10 service rates, from that in Experiment one.\textbf{ }That is, the
choice number $d=2$; the service rates of the $10$ servers are listed as
$\mu_{1}=1$, $\mu_{2}=2$, $\mu_{3}=6$, $\mu_{4}=8$, $\mu_{5}=10$, $\mu_{6}%
=16$, $\mu_{7}=17$, $\mu_{8}=18$, $\mu_{9}=25$ and $\mu_{10}=26$,
respectively; the probabilities of individual preference for the $10$ servers
are given by $g_{1}=0.10$, $g_{2}=0.20$, $g_{3}=0.30$, $g_{4}=0.05$,
$g_{5}=0.05$, $g_{6}=0.02$, $g_{7}=0.10$, $g_{8}=0.03$, $g_{9}=0.10$ and
$g_{10}=0.05$, respectively. We still simulate the expected queueing length
for each server by using the routine selection function%
\[
\Delta_{i}(x)=\frac{1+\frac{x_{i}}{\mu_{i}g_{i}}}{\underset{j=1}{\overset
{M}{\sum}}\left[  1+\frac{x_{j}}{\mu_{j}g_{j}}\right]  },\text{ \ }%
i=1,2,\ldots,10.
\]
The experimented results are shown in Table 2. It is seen from Tables 1 and 2
that the expected queue lengths of the $M$ servers decrease, as the service
rates of some servers increase.

\begin{table}[ptb]
\caption{The expected queue lengths in the 10 servers}%
\renewcommand{\arraystretch}{1.3}  \centering
\begin{tabular}
[c]{|c|c|c|}\hline
\bfseries {\bf Server number}  & \bfseries {\bf Expected queue lengths}  &
\\\hline
One & 0.3459 & \\\hline
Two & 0.1656 & \\\hline
Three & 0.0274 & \\\hline
Four & 0.0158 & \\\hline
Five & 0.0105 & \\\hline
Six & 0.0042 & \\\hline
Seven & 0.0038 & \\\hline
Eight & 0.0034 & \\\hline
Nine & 0.0018 & \\\hline
Ten & 0.0017 & \\\hline
\end{tabular}
\end{table}

\textbf{Experiment three:} Comparing with Experiments one and two, this
experiment takes more different parameters. We take that the choice number
$d=3$; the service rates of the $10$ servers are listed as $\mu_{1}=1$,
$\mu_{2}=3$, $\mu_{3}=3$, $\mu_{4}=6$, $\mu_{5}=6$, $\mu_{6}=6$, $\mu_{7}=6$,
$\mu_{8}=9$, $\mu_{9}=9$ and $\mu_{10}=15$, respectively; the probabilities of
individual preference for the $10$ servers are given by $g_{1}=0.05$,
$g_{2}=0.20$, $g_{3}=0.30$, $g_{4}=0.03$, $g_{5}=0.05$, $g_{6}=0.10$,
$g_{7}=0.10$, $g_{8}=0.05$, $g_{9}=0.02$ and $g_{10}=0.10$, respectively. We
simulate the expected queueing length for each server by using the routine
selection function%
\[
\Delta_{i}(x)=\frac{1+\frac{x_{i}}{\mu_{i}g_{i}}}{\underset{j=1}{\overset
{M}{\sum}}\left[  1+\frac{x_{j}}{\mu_{j}g_{j}}\right]  },\text{ \ }%
i=1,2,\ldots,10.
\]
The experimented results are shown in Table 3. It is seen from Tables 1, 2 and
3 that the expected queue lengths of the $M$ servers decrease largely, as the
choice number $d$ changes from $2$ to $3$. Therefore, ``the power of two
choices'' is still kept well in the study of supermarket models with different servers.

\begin{table}[ptb]
\caption{The expected queue lengths in the 10 servers}%
\renewcommand{\arraystretch}{1.3}  \centering
\begin{tabular}
[c]{|c|c|c|}\hline
\bfseries {\bf Server number}  & \bfseries {\bf Expected queue lengths}  &
\\\hline
One & 0.3447 & \\\hline
Two & 0.0580 & \\\hline
Three & 0.8598 & \\\hline
Four & 0.0265 & \\\hline
Five & 0.0265 & \\\hline
Six & 0.0266 & \\\hline
Seven & 0.0265 & \\\hline
Eight & 0.0126 & \\\hline
Nine & 0.0127 & \\\hline
Ten & 0.0048 & \\\hline
\end{tabular}
\end{table}

\section{Concluding Remarks}

In this paper, we provide a novel method for analyzing the supermarket model
with different servers through a multi-dimensional continuous-time Markov
reward process, and develop an event-driven technique both for computing the
mean of the random reward function in a finite time interval and for
calculating the mean of the discounted random reward function in an infinite
time interval. We indicate that the event-driven technique are useful in the
study of supermarket models with different servers, and more generally, in the
analysis of large-scale Markov reward processes. Notice that the supermarket
model with different servers is an important tool to set up some basic
relations between the system performance and the job routing rule, thus it can
also help to design reasonable architecture to improve the performance and to
balance the load in this supermarket model.

This paper provides a clear picture for how to use the event-driven technique
to analyze multi-dimensional continuous-time Markov reward processes, which
leads to performance analysis of the supermarket model with different servers.
We illustrate that this picture is organized as three key parts: (1)
Constructing a routine selection mechanism that depends on the queue lengths,
on the service rates, on the probability of individual preference and so
forth. (2) From the state jump points of the continuous-time Markov reward
process, we set up some segmented stochastic integrals of the random reward
function by means of an event-driven technique. Based on this, we compute the
mean of the random reward function in a finite time interval, and also
calculate the mean of the discounted random reward function in an infinite
time interval. Therefore, the results of this paper give new highlight on
understanding influence of the different servers on designing the routine
selection mechanism and on performance computation of more general supermarket
models. Along such a line, there are a number of interesting directions for
potential future research, for example,

\begin{itemize}
\item analyzing non-Poisson inputs such as, renewal processes; and discussing
non-exponential service time distributions, for example, general
distributions, matrix-exponential distributions and heavy-tailed distributions;

\item studying how to design a new routine selection mechanism with respect to
key random factors, such as, the least workload, and the subjective behavior
of customers;

\item developing effective algorithms both for computing the means of the
random reward functions and for solving the optimal problems in the study of
supermarket models with different servers; and

\item The event-driven technique is further developed for discussing the
sample paths of continuous-time Markov reward processes, thus the results
given in this paper may be very useful for performance simulation of more
general supermarket models with different servers.
\end{itemize}

Up to now, we believe that a larger gap exists when dealing with either
non-Poisson inputs or non-exponential service times in supermarket models with
different servers, because the event-driven technique needs be established for
being able to deal with more general Markov reward processes.

%\section*{Acknowledgements}

%The authors were supported by the National Natural Science Foundation of China
%under grant No. 71271187, 71471160 and 71471114, and the Fostering Plan of
%Innovation Team and Leading Talent in Hebei Universities under grant No. LJRC027.

\end{document}